\documentclass[12pt,showpacs,showkeys,amsmath,amssymb]{revtex4}
\usepackage{amsmath,amsfonts,amsthm,amscd,amssymb,latexsym}
\usepackage{bm}
\usepackage{dcolumn}
\usepackage{graphicx}
\usepackage{epstopdf}
\usepackage{color}
\usepackage{epsf}
\usepackage{epsfig}
\usepackage{graphicx, epic, eepic, color}

\def\d{\delta}

\def\@{\partial_}

\def\negenspace{\kern-1.1em}

\def\sqr#1#2{{\vcenter{\hrule height.#2pt\hbox{\vrule width.#2pt
height#1pt \kern#1pt \vrule width.#2pt}\hrule height.#2pt}}}
\def\square{\mathchoice\sqr64\sqr64\sqr{4.2}3\sqr{3.0}3}

\date{\today}

\begin{document}
\title{Nonlocal Gravity: The General Linear Approximation}

\author{B. Mashhoon}
\email{mashhoonb@missouri.edu}
\affiliation{Department of Physics and Astronomy,
University of Missouri, Columbia, Missouri 65211, USA}

\begin{abstract} 

The recent classical nonlocal generalization of Einstein's theory of gravitation is presented within the framework of general relativity via the introduction of a preferred frame field. The nonlocal generalization of Einstein's field equations is derived. The linear approximation of nonlocal gravity (NLG) is thoroughly examined and the solutions of the corresponding field equations are discussed. It is shown  that nonlocality, with a characteristic length scale of order 1 kpc, simulates dark matter in the linear regime while preserving causality. Light deflection in linearized nonlocal gravity is studied in connection with gravitational lensing; in particular, the propagation of light in the weak gravitational field of a uniformly moving  source is investigated. The astrophysical implications of the results are briefly mentioned. 

\end{abstract}



\maketitle

\section{Introduction}

The standard theory of relativity is based on a fundamental postulate of \emph{locality}. In extending Lorentz invariance to actual observers, which are all more or less \emph{accelerated}, a basic assumption is required regarding what accelerated observers would measure. The hypothesis that is adopted in the standard theory of relativity is that an arbitrary accelerated observer is pointwise inertial; therefore, Lorentz transformations can be applied point by point along the path of the accelerated observer to determine its measurements. This locality postulate is an essential ingredient of general relativity theory as well, since Einstein's heuristic principle of equivalence loses its significance if one does not know what accelerated observers would measure. In general relativity, an arbitrary observer in a gravitational field is locally (i.e., pointwise) inertial as a joint consequence of Einstein's principle of equivalence as well as the hypothesis of locality~\cite{Ei, Mas}. 

In classical physics, the value of a physical quantity $Q(t)$ at time $t$ is based on a certain measurement process that in general started \emph{before} time $t$. This circumstance becomes particularly significant when the \emph{acceleration} of the observer is taken into account, as a consequence of the existence of invariant acceleration scales in relativistic physics. Thinking of classical physics in terms of particles and waves, we note that the interaction of point particles and rays can be reduced to pointlike coincidences; however, one can show that wave properties cannot be measured instantaneously even with ideal measuring devices. The deviation from locality is thus expected to be proportional to $\lambdabar/{\mathcal L}$, where $\lambdabar$ is the reduced wavelength of the phenomenon under observation and $\mathcal{L}$ is the acceleration length of the observer. An observer fixed on the Earth, for instance, has translational and rotational acceleration lengths $c^{2}/|\mathbf{g}_{\oplus}|\approx 1$ light year and $c/|\boldsymbol{\omega}_{\oplus}|\approx 28$ astronomical units, respectively. Thus $\lambdabar/{\mathcal L}$ is generally rather small compared to unity; therefore, the hypothesis of locality is a good approximation in most situations of practical interest. Moreover, it has been shown by Bohr and Rosenfeld that the measurement of the classical electromagnetic field at a given time $t$ by ideal \emph{inertial} observers involves a certain spacetime average over past events~\cite{BR}. This observation acquires particular significance for \emph{accelerated} observers in Minkowski spacetime due to the existence of invariant acceleration scales~\cite{BM}. Thus the application of the hypothesis of locality to a basic field such as the electromagnetic field is only a first approximation, akin to the impulse approximation of the quantum scattering theory~\cite{Mash}. 

To go beyond the hypothesis of locality, one must include an average over the past world line of the observer with a weight function that is characteristic of the observer's acceleration. In this way, a nonlocal special relativity theory has been developed in which nonlocality appears as the \emph{memory} of  past acceleration~\cite{M1}. Thus the measured electromagnetic field consists of the local result plus an integral over the past that is linear in the field and contains an acceleration-dependent kernel~\cite{MHM}. The electromagnetic field is \emph{local}, but satisfies integro-differential equations reminiscent of Maxwell's original equations with \emph{nonlocal constitutive relations}~\cite{M2}. How can this approach be extended to a nonlocal general relativity theory~\cite{M3}? It turns out that general relativity (GR) has an equivalent tetrad formulation (GR$_{||}$) that is amenable to  nonlocal generalization via a  causal constitutive kernel~\cite{NL1, NL2, NL3, NL4, NL5, NL6, NL7, NL8}. Such a nonlocal generalization of GR can simulate dark matter. The fundamental length scale associated with nonlocal gravity (NLG) is a galactic length of the order of 1 kpc; therefore, nonlocality can be neglected on scales that are much smaller than 1 kpc. It appears that the nonlocal aspect of gravity could indeed be responsible for the observational data in astrophysics and cosmology that have been interpreted thus far in terms of dark matter; that is, what is now considered dark matter may in fact be the manifestation of the nonlocal component of the gravitational interaction~\cite{NL8}. This circumstance provides the motivation to study the theoretical basis of nonlocal gravity (NLG) further and develop its consequences. 

In previous work on NLG~\cite{NL1, NL2, NL3, NL4, NL5, NL6, NL7, NL8}, GR$_{||}$, the teleparallel equivalent of GR, has been described within the framework of gauge theories of gravitation, since GR$_{||}$ is the gauge theory of the Abelian group of spacetime translations~\cite{BH, AP, Ma}. Alternatively, it is possible to formulate the theory within the standard framework of GR supplemented with a latticework of preferred frames. For the sake of completeness, we adopt in the present paper, the latter formulation that is much closer to the spirit of GR~\cite{Ma}. That is, nonlocal gravity has been primarily described thus far in terms of local frames in Weitzenb\"ock spacetime~\cite{NL1, NL2, NL3, NL4, NL5, NL6, NL7, NL8}.  We choose a complementary approach in this paper and formulate nonlocal gravity anew in such a way as to preserve the main physical results of the theory~\cite{NL1, NL2, NL3, NL4, NL5, NL6, NL7, NL8} and avoid inconsistencies, as explained in detail in the following sections. Furthermore, the matter energy-momentum tensor $T_{\mu \nu}$, as employed in previous work on nonlocal gravity~\cite{NL1, NL2, NL3, NL4, NL5, NL6, NL7, NL8}, has not always been assumed to be  symmetric in general. In the present work, however, $T_{\mu \nu}$ is the symmetric energy-momentum tensor, exactly as in GR.   

In section II, we introduce the Weitzenb\"ock connection and concisely develop the essential elements of nonlocal gravity  in an extended general relativistic framework. In particular, the field equations of nonlocal gravity are written as nonlocally modified Einstein's equations. In sections III and IV, the general linear approximation of nonlocal gravity is developed in a consistent manner and applied in section V to the determination of the gravitational field of an isolated stationary source. Such a source is assumed to be in uniform translational motion in section VI, which is devoted to the problem of propagation of light rays in the field of the moving source. Section VII contains a brief discussion of our results. 

\section{Field Equations of Nonlocal Gravity}

\subsection{One Metric with Two Connections}

Einstein's local principle of equivalence has a natural geometric formulation in terms of a spacetime manifold with a Riemannian metric tensor $g_{\mu \nu}(x)$ such that test particles follow timelike geodesics 
\begin{equation}\label{II1}
\frac{d^2 x^{\mu}}{d\tau^2} + {^0}\Gamma^\mu_{\alpha \beta}\, \frac{dx^\alpha}{d\tau} \frac{dx^\beta}{d\tau} = 0\,
\end{equation}
and rays of radiation follow the corresponding null geodesics of spacetime~\cite{Ei}. Here, $x$ represents an event in spacetime with coordinates  $x^\mu=(ct, x^i)$, $\tau$ is the proper time along the world line  and (${^0}\Gamma^\mu_{\alpha \beta}$) represents the Levi-Civita connection given by the symmetric Christoffel symbols
\begin{equation}\label{II2}
{^0}\Gamma^\mu_{\alpha \beta}= \frac{1}{2} g^{\mu \nu} (g_{\nu \alpha,\beta}+g_{\nu \beta,\alpha}-g_{\alpha \beta,\nu})\,.
\end{equation}
This torsion-free connection has Riemannian curvature,
\begin{equation}\label{II2a}
^{0}R^{\alpha}{}_{\mu \beta \nu}=\partial_{\beta}\, {^0} \Gamma^\alpha_{\nu \mu} -\partial_{\nu}\, {^0} \Gamma^\alpha_{\beta \mu}+\,^{0}\Gamma^{\alpha}_{\beta \gamma}\, ^{0}\Gamma^{\gamma}_{\nu \mu}-\,^{0}\Gamma^\alpha_{\nu \gamma}\, ^{0}\Gamma^\gamma_{\beta \mu}\,,
\end{equation}
in terms of which one can develop a natural generalization of Poisson's equation of Newtonian gravity. Hence, we have the gravitational field equations~\cite{Ei} 
\begin{equation}\label{II3}
 {^0}R_{\mu \nu}-\frac{1}{2} g_{\mu \nu}\,{^0}R+ \Lambda\, g_{\mu \nu}=\kappa\,T_{\mu \nu}\,,
 \end{equation}
where ${^0}R_{\mu \nu}={^0}R^{\alpha}{}_{\mu \alpha \nu}$ represents Ricci curvature and ${^0}R= g^{\mu \nu}~{^0}R_{\mu \nu}$ represents scalar curvature. Moreover, the matter energy-momentum tensor is \emph{symmetric} and given by $T_{\mu \nu}$, $\Lambda$ is the cosmological constant and $\kappa:=8 \pi G/c^4$. The Einstein equations can be derived from an action principle, where the gravitational Lagrangian is given by $L_g =c ^3({^0}R-2\Lambda)/(16 \pi G)$. The gravitational field is identified with the Riemannian curvature tensor ${^0}R_{\alpha \beta \gamma \delta}$; in its complete absence, there is no gravity and we are back in the Minkowski spacetime of special relativity~\cite{Sy}. 

Observers in spacetime are endowed with an orthonormal tetrad frame $\lambda^\mu{}_{\hat {\alpha}}(x)$ such that $\lambda^\mu{}_{\hat {0}}$ is the observer's unit temporal direction, $\lambda^\mu{}_{\hat {i}}$\,, $i=1, 2, 3,$ form its spatial frame and 
\begin{equation}\label{II4}
 g_{\mu \nu}(x) \, \lambda^\mu{}_{\hat {\alpha}}(x)~ \lambda^\nu{}_{\hat {\beta}}(x)= \eta_{\hat {\alpha} \hat  {\beta}}\,.
\end{equation}
The 16 components of the tetrad frame are subject to 10 orthonormality relations~\eqref{II4}. Let us recall that in GR, the metric tensor $g_{\mu \nu}$ carries the 10  gravitational degrees of freedom. The remaining 6 degrees of freedom, which are elements of the local Lorentz group, specify  the observer's instantaneous velocity and the 3 Euler angles that define the orientation of its spatial frame with respect to a background reference system.  

The \emph{local} measurement of physical quantities by an observer generally involves the projection of relevant tensor fields on its tetrad frame.
Thus the spacetime interval $ds$ can be written as 
\begin{equation}\label{II5}
ds^2 = g_{\mu \nu}~dx^\mu \, dx^\nu =\eta_{\hat {\alpha} \hat {\beta}}~dx^{\hat {\alpha}} \, dx^{\hat {\beta}}\,,
\end{equation}
where $dx^\mu=\lambda^\mu{}_{\hat {\alpha}}~dx^{\hat {\alpha}}$. In our convention, the Minkowski metric tensor  $\eta_{\alpha \beta}$ is given by diag$(-1,1,1,1)$. Moreover, Latin indices run from 1 to 3, unless specified otherwise, while Greek indices run from 0 to 3. The \emph{hatted} indices (e.g., $\hat {\alpha}$, $\hat {i}$, etc.) refer to \emph{anholonomic} tetrad---that is, local Lorentz---indices, while ordinary indices (e.g., $\alpha$, $i$, etc.) refer to general \emph{holonomic} spacetime indices. As is evident from Eq.~\eqref{II5}, the tetrad connects (holonomic) spacetime quantities to (anholonomic) local Lorentz quantities. \emph{In keeping with the spirit of GR, we work in this paper essentially with holonomic systems and the corresponding spacetime coordinates are assumed to be admissible.} Holonomic and anholonomic indices are raised and lowered by means of metric tensors $g_{\mu \nu}(x)$ and $\eta_{\hat {\alpha} \hat {\beta}}$, respectively. To change an anholonomic index of a tensor into a holonomic index or vice versa, we simply project the tensor onto the corresponding tetrad frame. We use units such that $c=1$, unless otherwise specified.

Of all possible smooth orthonormal tetrad frame fields that can be defined on the Riemannian spacetime manifold, let us choose one, namely, $e^\mu{}_{\hat{\alpha}}(x)$. This will be our \emph{preferred} tetrad field. Indeed, any such smooth frame field will do; however, this basic degeneracy will be eventually removed via the introduction of nonlocality into the theory.  Let us now use our preferred frame to define a second connection~\cite{We}
\begin{equation}\label{II6}
\Gamma^\mu_{\alpha \beta}=e^\mu{}_{\hat{\rho}}~\partial_\alpha\,e_\beta{}^{\hat{\rho}}\,.
\end{equation}
One can directly verify that this nonsymmetric \emph{Weitzenb\"ock connection} is indeed curvature-free. It follows from Eq.~\eqref{II6} that 
\begin{equation}\label{II7}
\nabla_\nu\,e_\mu{}^{\hat{\alpha}}=0\,,
\end{equation}
where $\nabla$ denotes covariant differentiation with respect to the Weitzenb\"ock connection. Therefore,  $\nabla_\nu\, g_{\alpha \beta}=0$ due to the orthonormality relation
\begin{equation}\label{II8}
g_{\mu \nu} = e_\mu{}^{\hat{\alpha}}~ e_\nu{}^{\hat{\beta}}~ \eta_{\hat{\alpha} \hat{\beta}}\,,
\end{equation}
so that the Weitzenb\"ock connection is compatible with the spacetime metric. Moreover, the new connection renders spacetime a parallelizable manifold, since we have everywhere access to our preferred frame field $e^\mu{}_{\hat{\alpha}}$, a smooth global latticework of parallel tetrad frames. This framework is known as teleparallelism, due to the distant parallelism of the preferred tetrad frames via the Weitzenb\"ock connection. That is, distant vectors can be considered parallel if they have the same local components relative to the preferred tetrad frame field. 

We have thus two connections that are both compatible with our Riemannian metric. The difference between two connections on the same manifold is always a tensor; therefore, we have two associated tensor fields, namely, the \emph{torsion} tensor
\begin{equation}\label{II9}
 C_{\mu \nu}{}^{\alpha}=\Gamma^{\alpha}_{\mu \nu}-\Gamma^{\alpha}_{\nu \mu}=e^\alpha{}_{\hat{\beta}}\Big(\partial_{\mu}e_{\nu}{}^{\hat{\beta}}-\partial_{\nu}e_{\mu}{}^{\hat{\beta}}\Big)\,,
\end{equation}
and the \emph{contorsion} tensor
\begin{equation}\label{II10}
K_{\mu \nu}{}^\alpha= {^0} \Gamma^\alpha_{\mu \nu} - \Gamma^\alpha_{\mu \nu}\,.
\end{equation}
From $\nabla_\gamma\,g_{\alpha \beta}=0$, we have
\begin{equation}\label{II11}
 g_{\alpha \beta , \gamma}= \Gamma^\mu_{\gamma \alpha}\, g_{\mu \beta} + \Gamma^\mu_{\gamma \beta}\, g_{\mu \alpha}\,.
\end{equation}
Substituting this relation in Eq.~\eqref{II2}, we find the relation between torsion and contorsion, namely,
\begin{equation}\label{II12}
K_{\mu \nu}{}^\alpha = \frac{1}{2}\, g^{\alpha \beta} (C_{\mu \beta \nu}+C_{\nu \beta \mu}-C_{\mu \nu \beta})\,.
\end{equation}
While the torsion tensor is antisymmetric in its first two indices, the contorsion tensor is antisymmetric in its last two indices.

We have identified the gravitational field with the Riemann curvature tensor ${^0}R_{\mu \nu \rho \sigma}$ from the standpoint of the Levi-Civita connection. From the standpoint of the Weitzenb\"ock connection, however, the gravitational field would naturally be identified with the torsion tensor $C_{\mu \nu \rho}$. It can be shown that these notions are indeed compatible~\cite{M0}. To see briefly how this can come about, let us consider the torsion tensor in the form
\begin{equation}\label{II13}
 C_{\mu \nu}{}^{\hat{\alpha}}=e_\rho{}^{\hat{\alpha}}C_{\mu \nu}{}^{\rho}= \partial_{\mu}e_{\nu}{}^{\hat{\alpha}}-\partial_{\nu}e_{\mu}{}^{\hat{\alpha}}\,.
\end{equation}
For each ${\hat{\alpha}}={\hat{0}}, {\hat{1}}, {\hat{2}}, {\hat{3}}$, we have in Eq.~\eqref{II13} an analog of the electromagnetic field tensor defined in terms of the vector potential $e_{\mu}{}^{\hat{\alpha}}$. The field completely vanishes if the potential is a pure gauge; that is, if there are functions $X^{\hat{\alpha}}$ such that $e_{\mu}{}^{\hat{\alpha}}=\partial_\mu X^{\hat{\alpha}}$. It then follows via Eq.~\eqref{II8} that we are indeed in Minkowski spacetime and ${^0}R_{\mu \nu \rho \sigma}=0$. Conversely, in a gravitational field with ${^0}R_{\mu \nu \rho \sigma} \ne 0$, the torsion tensor is necessarily nonzero. It is therefore natural to express Einstein's field equations in terms of the torsion tensor. It is not surprising that the result will turn out to be reminiscent of Maxwell's equations.  This way of describing the gravitational field, namely, GR$_{||}$, the teleparallel equivalent of GR, turns out to be crucial for a proper nonlocal generalization of GR~\cite{NL1, NL2}.  Appendix A contains a set of formulas involving torsion and contorsion that should be useful in writing the field equations in terms of torsion. 

\subsection{GR$_{||}$}

We can now combine Eqs.~\eqref{II10} and~\eqref{II12} in order to express the Levi-Civita connection in terms of the Weitzenb\"ock connection and its torsion tensor. Substituting the result in the Riemann tensor~\eqref{II2a} and taking the appropriate trace,  we find that the Ricci tensor, ${^0}R_{\mu \nu}={^0}R^{\alpha}{}_{\mu \alpha \nu}$, is given by 
\begin{eqnarray}\label{II14}
 {^0}R_{\mu \nu}=\frac{1}{\sqrt{-g}}\,\frac{\partial}{\partial x^\alpha}\, \Big(\sqrt{-g}\,K_{\nu \mu}{}^{\alpha}\Big)+\frac{\partial C_\mu}{\partial x^\nu}-C_\alpha \Gamma^\alpha_{\nu \mu}\nonumber \\
-(\Gamma^\alpha_{\nu \beta}+ K_{\nu \beta}{}^\alpha)K_{\alpha \mu}{}^{\beta}-\Gamma^{\alpha}_{\beta \mu}K_{\nu \alpha}{}^{\beta}\,.
\end{eqnarray}      
Here $g:=\det(g_{\mu \nu})$, $\sqrt{-g}=\det(e_{\mu}{}^{\hat{\alpha}})$ and $C_\mu$ is the \emph{torsion vector}, which is the trace of the torsion tensor; that is, 
\begin{equation}\label{II15}
C_\mu :=C^{\alpha}{}_{\mu \alpha} = - C_{\mu}{}^{\alpha}{}_{\alpha}\,. 
\end{equation}

To express the gravitational field equations in terms of our preferred frame field $e_{\mu}{}^{\hat{\alpha}}$ and its torsion tensor, we first note that the scalar curvature can be obtained from the trace of the Ricci tensor, namely,  
\begin{equation}\label{II16}
{^0}R=-\frac{1}{2}
\mathfrak{C}_{\alpha \beta \gamma}C^{\alpha \beta \gamma}+\frac{2}{\sqrt{-g}}\,\frac{\partial}{\partial x^\delta}\, \Big(\sqrt{-g}\,C^{\delta}\Big)\,,
\end{equation}
where $\mathfrak{C}_{\alpha \beta \gamma}$ is the \emph{auxiliary torsion tensor} that is also antisymmetric in its first two indices and is defined by
\begin{equation}\label{II17}
\mathfrak{C}_{\alpha \beta \gamma} :=C_\alpha\, g_{\beta \gamma} - C_\beta \,g_{\alpha \gamma}+K_{\gamma \alpha \beta}\,.
\end{equation}

Let us briefly digress here and mention that the Lagrangian for GR$_{||}$ contains only the first term on the right-hand side of Eq.~\eqref{II16}, as the second term turns into a surface term in the action. Moreover, 
\begin{equation}\label{II18}
\mathfrak{C}_{\alpha \beta \gamma}C^{\alpha \beta \gamma}= \frac{1}{2} I_1 + I_2 -2 I_3\,,
\end{equation}
where
\begin{equation}\label{II19}
I_1=C_{\alpha \beta \gamma}C^{\alpha \beta \gamma}, \quad I_2=C_{\alpha \beta \gamma}C^{\gamma \beta \alpha}, \quad I_3=C_\alpha C^\alpha\,
\end{equation}
are the three independent algebraic (Weitzenb\"ock) invariants of the torsion tensor. 

We now introduce a second auxiliary field strength  ${\cal H}^{\mu \nu}{}_{\rho}=-{\cal H}^{\nu \mu}{}_{\rho}$ defined by
\begin{equation}\label{II20}
{\cal H}_{\mu \nu \rho}:= \frac{\sqrt{-g}}{\kappa}\,\mathfrak{C}_{\mu \nu \rho}\,.
\end{equation}
It proves useful for our present purposes to express the Einstein tensor as  ${^0}G_{\mu \nu}={^0}R_{\nu \mu}-\frac{1}{2}g_{\mu \nu}\,{^0}R$, where the indices on the symmetric Ricci tensor have been switched in order to get from Eqs.~\eqref{II14} and~\eqref{II16} the Einstein tensor in the form
\begin{eqnarray}\label{II21}
 {^0}G_{\mu \nu}=\frac{\kappa}{\sqrt{-g}}\Big[g_{\nu \alpha}\, e_\mu{}^{\hat{\gamma}}\,\frac{\partial}{\partial x^\beta}\,{\cal H}^{\alpha \beta}{}_{\hat{\gamma}}
-\Big({\cal H}_{\nu \rho \sigma}C_{\mu}{}^{\rho \sigma}
-\frac{1}{4}g_{\nu \mu}\,{\cal H}_{\alpha \beta \gamma}C^{\alpha \beta \gamma}\Big) \Big]\,.
\end{eqnarray}      
Thus the Einstein field equations~\eqref{II3} can be written within the GR$_{||}$ framework in the  Maxwellian form
\begin{equation}\label{II22}
 \frac{\partial}{\partial x^\nu}\,{\cal H}^{\mu \nu}{}_{\hat{\alpha}}+\frac{\sqrt{-g}}{\kappa}\,\Lambda\,e^\mu{}_{\hat{\alpha}} =\sqrt{-g}~(T_{\hat{\alpha}}{}^\mu + E_{\hat{\alpha}}{}^\mu)\,,
\end{equation}
where $E_{\mu \nu}$ is now the trace-free energy-momentum tensor of the gravitational field defined by
\begin{equation}\label{II23}
\sqrt{-g}~ E_{\hat{\alpha}}{}^\mu:=C_{{\hat{\alpha}} \rho \sigma} {\cal H}^{\mu \rho \sigma}-\frac 14  e^\mu{}_{\hat{\alpha}}~C_{ \nu \rho \sigma}{\cal H}^{\nu \rho \sigma}\,.
\end{equation}
It follows from Eq.~\eqref{II22} and the antisymmetry of ${\cal H}^{\mu \nu}{}_{\hat{\alpha}}$ in its first two indices that 
\begin{equation}\label{II24}
\frac{\partial}{\partial x^\mu}\,\Big[\sqrt{-g}~(T_{\hat{\alpha}}{}^\mu-\frac{\Lambda}{\kappa}\,e^\mu{}_{\hat{\alpha}} + E_{\hat{\alpha}}{}^\mu)\Big]=0\,,
 \end{equation}
which expresses the conservation law of total energy-momentum tensor in GR$_{||}$, consisting of contributions due to matter, the cosmological constant and the gravitational field, respectively. We emphasize that the procedure we have followed would work for any smooth tetrad field that we may adopt as our preferred frame. This is related to the invariance of Einstein's theory under the \emph{local} Lorentz group. That is, Eq.~\eqref{II22} ultimately depends only upon the metric tensor $g_{\mu \nu}$; therefore, this teleparallel formulation involves a 6-fold degeneracy at each event in spacetime. 

The tetrad formulation of GR has a long history---see Refs.~\cite{BH, AP, Ma} and the references cited therein. Indeed,  M{\o}ller first pointed out that the problem of gravitational energy in GR has a solution in the tetrad framework~\cite{Mo, PP}. An excellent review of the approach to GR$_{||}$ that we have adopted in the present paper has been given by Maluf~\cite{Ma}, which should be consulted for further developments of GR$_{||}$. This concludes our brief presentation of the salient features of GR$_{||}$, the 
teleparallel equivalent of GR. 

\subsection{Nonlocal GR$_{||}$}

In his successful approach to GR, Einstein interpreted the experimentally well-established principle of equivalence of inertial and gravitational masses to mean that there is an intimate connection between \emph{inertia} and \emph{gravitation}~\cite{Ei}. This notion eventually led to Einstein's extremely \emph{local} principle of equivalence and GR. Following Einstein, we wish to employ the general connection between inertia and gravitation as a guiding principle to render GR (or, equivalently, GR$_{||}$) \emph{nonlocal} in just the same way that accelerated observers in Minkowski spacetime are nonlocal. In field measurements of accelerated observers, the memory of past acceleration appears as an integral over the past that is linear in the field. To implement the same idea in the theory of gravitation, we note that Einstein's field equations, represented by Eq.~\eqref{II22} in our tetrad framework, have the general form of Maxwell's original field equations with the \emph{local} constitutive relation~\eqref{II20}. To render GR$_{||}$ nonlocal, we simply replace the local constitutive relation~\eqref{II20} with a nonlocal one given by  
\begin{equation}\label{II25}
{\cal H}_{\mu \nu \rho}:= \frac{\sqrt{-g}}{\kappa}(\mathfrak{C}_{\mu \nu \rho}+ N_{\mu \nu \rho})\,,
\end{equation}
where $N_{\mu \nu \rho}$ is a tensor involving an average of the gravitational field---that is, torsion---over past events. We emphasize that in order to preserve the invariance of the theory under arbitrary coordinate transformations,  $N_{\mu \nu \rho}$ and hence the resulting \emph{nonlocal} auxiliary field strength ${\cal H}_{\mu \nu \rho}$ should be antisymmetric in their first two indices. The simplest expression for the nonlocality tensor $N_{\mu \nu \rho}$ would involve a \emph{scalar} kernel; that is, 
\begin{eqnarray}\label{II26}
N_{\mu \nu \rho} = - \int \Omega_{\mu \mu'} \Omega_{\nu \nu'} \Omega_{\rho \rho'}\, {\cal K}(x, x')\,X^{\mu' \nu' \rho'}(x') \sqrt{-g(x')}\, d^4x' \,,
\end{eqnarray}
where ${\cal K}$ is the scalar \emph{causal} kernel of the nonlocal theory~\cite{NL1, NL2, NL3, NL4, NL5, NL6, NL7, NL8} and  $X_{\mu \nu \rho}(x)$ is a tensor that is antisymmetric in its first two indices and  involves a linear combination of the components of the torsion tensor. We note that there is no physical connection between kernel ${\cal K}$ and the nonlocal kernel of accelerated observers in Minkowski spacetime due to the extreme locality of Einstein's principle of equivalence. In Eq.~\eqref{II26}, $\Omega(x, x')$ is Synge's \emph{world function}~\cite{Sy}, which involves a unique future-directed timelike or null geodesic of $g_{\mu \nu}$  that connects event $x'$ to event $x$ and the square of its proper length is 2\,$\Omega$.  Moreover, indices $\mu', \nu', \rho',...$ refer to event $x'$, while indices $\mu, \nu, \rho, ...$ refer to event $x$.  We define
\begin{equation}\label{II27}
\Omega_{\mu}(x, x'):=\frac{\partial \Omega}{\partial x^{\mu}}, \qquad \Omega_{\mu'}(x, x'):=\frac{\partial \Omega}{\partial x'^{\mu'}}\,.
\end{equation}
It can be shown that covariant derivatives at $x$ and $x'$ commute for any bitensor~\cite{Sy}. Thus $\Omega_{\mu \mu'}(x, x')=\Omega_{\mu' \mu}(x, x')$ is a dimensionless bitensor such that 
\begin{equation}\label{II28}
\lim_{x' \to x} \Omega_{\mu \mu'}(x, x')=-g_{\mu \mu'}(x)\,.
\end{equation}

Let us now consider the field equations of nonlocal GR$_{||}$ with a general $X_{\mu \nu \rho}=-X_{\nu \mu \rho}$. The field equations of nonlocal gravity (NLG), namely, Eqs.~\eqref{II22}--\eqref{II23} together with the nonlocal constitutive relation~\eqref{II25} can be expressed explicitly by substituting Eq.~\eqref{II25} in Eqs.~\eqref{II22} and~\eqref{II23}. Thus, we have 
\begin{equation}\label{II28b}
 \frac{\partial}{\partial x^\nu}\,\Big[\frac{\sqrt{-g}}{\kappa}\,(\mathfrak{C}^{\mu \nu}{}_{\hat{\alpha}}+N^{\mu \nu}{}_{\hat{\alpha}})\Big]+\frac{\sqrt{-g}}{\kappa}\,\Lambda\,e^\mu{}_{\hat{\alpha}} =\sqrt{-g}~(T_{\hat{\alpha}}{}^\mu + E_{\hat{\alpha}}{}^\mu)\,,
\end{equation}
where $E_{\hat{\alpha}}{}^\mu$ is now given by 
\begin{equation}\label{II28c}
\kappa~ E_{\hat{\alpha}}{}^\mu:=C_{{\hat{\alpha}} \rho \sigma} (\mathfrak{C}^{\mu \rho \sigma}+N^{\mu \rho \sigma})-\frac 14  e^\mu{}_{\hat{\alpha}}~C_{ \nu \rho \sigma}(\mathfrak{C}^{\nu \rho \sigma}+N^{\nu \rho \sigma})\,.
\end{equation}
With this $E_{\hat{\alpha}}{}^\mu$, the total energy-momentum conservation law~\eqref{II24} is satisfied; that is, in nonlocal gravity, energy-momentum  conservation is represented by a simple generalization of  Eq.~\eqref{II24} of GR$_{||}$, where $E_{\hat{\alpha}}{}^\mu$ is given by Eq.~\eqref{II28c}.

It is possible to express the nonlocal gravitational field equations as modified Einstein's equations.  To this end, we separate out in Eq.~\eqref{II28b} the partial derivative term involving $(\sqrt{-g}/\kappa)\,\mathfrak{C}^{\mu \nu}{}_{\hat{\alpha}}$ and insert it into the expression~\eqref{II21} for the Einstein tensor $ {^0}G_{\mu \nu}$ to get the \emph{nonlocal generalization of Einstein's field equations}, namely, 
\begin{equation}\label{II29}
 {^0}G_{\mu \nu}+{\cal N}_{\mu \nu}=\kappa\, T_{\mu \nu}- \Lambda\, g_{\mu \nu}+Q_{\mu \nu}\,.
 \end{equation}
Here,  ${\cal N}_{\mu \nu}$ defined by
\begin{equation}\label{II29a}
{\cal N}_{\mu \nu}:=g_{\nu \alpha}\, e_\mu{}^{\hat{\gamma}}\,\frac{1}{\sqrt{-g}}\,\frac{\partial}{\partial x^\beta}\,\Big(\sqrt{-g}\,N^{\alpha \beta}{}_{\hat{\gamma}}\Big)
\end{equation}
is a proper tensor, since $N_{\alpha \beta \gamma}=-N_{\beta \alpha \gamma}$ by assumption; moreover, $Q_{\mu \nu}$ is a traceless tensor given by
\begin{equation}\label{II30}
Q_{\mu \nu}:=C_{\mu \rho \sigma} N_{\nu}{}^{\rho \sigma}-\frac 14\, g_{\mu \nu}\,C_{ \delta \rho \sigma}N^{\delta \rho \sigma}\,.
\end{equation}
It is clear that Einstein's gravitational field equations are recovered when the nonlocal kernel vanishes, ${\cal K}=0$, and hence $N_{\mu \nu \rho}=0$. In GR, the 10 components of the metric tensor $g_{\mu \nu}$ can  be determined, in principle, from the 10 gravitational field equations. Here, however, the 16 components of the preferred observers' frame field $e^\mu{}_{\hat{\alpha}}$ can be obtained, in principle, from the 16 gravitational field equations~\eqref{II29}--\eqref{II30} of nonlocal general relativity. That is, nonlocality removes the essential degeneracy of GR$_{||}$; moreover, as expected, nonlocal gravity is invariant under the \emph{global} Lorentz group. The integro-differential field equations of nonlocal gravity in general contain Fredholm integral relations that, whenever causal kernels are involved, turn into Volterra integral relations~\cite{WVL, Tr}.

To compare and contrast further the field equations of nonlocal gravity with the Einstein field equations of GR, one can separate out Eq.~\eqref{II29} into its symmetric and antisymmetric components.  In this way, we get the 10 nonlocally modified Einstein equations given by 
\begin{equation}\label{II31}
 {^0}G_{\mu \nu}+{\cal N}_{(\mu \nu)}=\kappa\, T_{\mu \nu}- \Lambda\, g_{\mu \nu}+Q_{(\mu \nu)}\, \end{equation}
as well as the 6 integral constraint equations involving the nonlocality tensor $N_{\mu \nu \rho}$, namely, 
\begin{equation}\label{II32}
{\cal N}_{[\mu \nu]}=Q_{[\mu \nu]}=\frac{1}{2}\Big(C_{\mu \rho \sigma} N_{\nu}{}^{\rho \sigma}-C_{\nu \rho \sigma} N_{\mu}{}^{\rho \sigma}\Big)\,,
\end{equation}
that are dominated by averaging over past events and vanish for ${\cal K}=0$. The energy-momentum tensor is symmetric in this paper; therefore, there is no contribution from $T_{[\mu \nu]}=0$ to Eq.~\eqref{II32}. This point brings out the main difference between the present work and previous papers on nonlocal gravity~\cite{NL1, NL2, NL3, NL4, NL5, NL6, NL7, NL8}, in which $T_{\mu \nu}$ was not assumed to be symmetric from the outset.  
Let us recall here that these 16 field equations are required to determine the 16 components of 
$e^{\mu}{}_{\hat{\alpha}}(x)$, of which 10 are fixed by the spacetime metric $g_{\mu \nu}$ via orthonormality and the other 6 are Lorentz degrees of freedom (i.e., boosts and rotations). This division is reflected in Eqs.~\eqref{II31} and~\eqref{II32}, respectively. The general mathematical investigation of the existence and uniqueness of the solutions of the integro-differential Eq.~\eqref{II31} with integral constraints~\eqref{II32} is beyond the scope of the present paper. 

It is worthwhile to emphasize again the close analogy between this construction of nonlocal gravity and the nonlocal electrodynamics of media. Maxwell's equations in a medium in an inertial frame can be expressed in terms of the field tensors $F_{\mu\nu} \mapsto (\mathbf{E}, \mathbf{B})$ and $H_{\mu\nu} \mapsto (\mathbf{D}, \mathbf{H})$ as 
\begin{equation}\label{II33}
F_{[\mu \nu , \rho]}=0\,,  \qquad   \partial_{\nu}\,H^{\mu \nu}=\frac{4\pi}{c}\, J^{\mu}\,,
\end{equation}
where $J^\mu$ is the total current 4-vector associated with \emph{free} electric charges. To complete the theory, a constitutive relation between $F_{\mu\nu}$ and $H_{\mu\nu}$ is required. If we impose the local relation $H_{\mu\nu}=F_{\mu\nu}$, we recover Maxwell's equations in vacuum. However, in a medium the constitutive relation is in general nonlocal~\cite{Ja, CGAE}, thus leading to the nonlocal electrodynamics of media. In the gravitational case, on the other hand, Einstein's field equations have been expressed within the teleparallelism framework with the local constitutive relation~\eqref{II20}  in a form analogous to Maxwell's equations in vacuum. We have then extended this relation to a nonlocal one via our ansatz~\eqref{II25}, which has therefore resulted in a simple nonlocal extension of Einstein's theory of gravitation.  Let us note here that the constitutive ansatz~\eqref{II25} involves a \emph{linear} nonlocal relation between the two field strengths involving ${\cal H}_{\mu \nu \rho}$ and $\mathfrak{C}_{\mu \nu \rho}$; however, as in electrodynamics~\cite{Ja, CGAE}, such a nonlocal  relation could well become nonlinear  when the field strengths are sufficiently high. We will not have occasion here to discuss such nonlinearities, since at this early stage in the development of NLG the relation between $X_{\mu \nu \rho}$ and torsion is assumed to be linear for the sake of simplicity. 

In electrodynamics, the local constitutive relation between $H_{\mu \nu}$ and $F_{\mu \nu}$, considered as 6-vectors, can be described via a $6 \times 6$ matrix. One can similarly envision the local linear  relationship between $X_{\mu \nu \rho}=-X_{\nu \mu \rho}$ and $\mathfrak{C}_{\mu \nu \rho}$ in Eq.~\eqref{II26} in a rather general context. The general case is beyond the scope of the present work; instead, we limit our considerations to a few simple observations regarding such relations here and in Appendix B. We assume that the constitutive relations are given up to constant overall multiplicative factors, since these could be absorbed in the corresponding scalar kernels. Previous work on NLG has been based on the simplest constitutive relation, namely, $X_{\mu \nu \rho}= \mathfrak{C}_{\mu \nu \rho}$~\cite{NL1, NL2, NL3, NL4, NL5, NL6, NL7, NL8}. However, in contrast to previous work~\cite{NL1, NL2, NL3, NL4, NL5, NL6, NL7, NL8}, we assume here from the outset that $T_{\mu \nu}$ is symmetric. Then, as we show in detail in Appendix B, $X_{\mu \nu \rho}= \mathfrak{C}_{\mu \nu \rho}$ is in general untenable in linearized NLG. We must therefore explore other options. Of the various possibilities of the general form $X_{\mu \nu \rho}= \mathfrak{C}_{\mu \nu \rho} +A_{\mu \nu \rho}$ that we have considered, additions $A_{\mu \nu \rho}$ of the forms $C_\mu\, g_{\nu \rho}-C_\nu\, g_{\mu \rho}$ and $\check{C}_\mu\, g_{\nu \rho}-\check{C}_\nu\, g_{\mu \rho}$ have been found to be tenable in the linear approximation. Here $\check{C}^\mu$ is the torsion pseudovector
\begin{equation}\label{II34}
\check{C}_\mu=\frac{1}{3!} C^{\alpha \beta \gamma}\,E_{\alpha \beta \gamma \mu}\,,
\end{equation}
which is the dual of the torsion tensor, see Appendix A. The 16 gravitational potentials of \emph{linearized} NLG can be divided into 10 \emph{metric} potentials and 6 \emph{tetrad} potentials. It turns out that the torsion vector contains both metric and tetrad potentials, while the torsion pseudovector contains only the tetrad potentials.  The latter leads to much simplification; hence, in this paper, we tentatively choose the local constitutive relation of NLG to be
\begin{equation}\label{II35}
X_{\mu \nu \rho}= \mathfrak{C}_{\mu \nu \rho}+ p\,(\check{C}_\mu\, g_{\nu \rho}-\check{C}_\nu\, g_{\mu \rho})\,,
\end{equation}
where $p\ne 0$ is a constant dimensionless parameter. We emphasize again that this is different from previous work on nonlocal gravity, where $X_{\mu \nu \rho}= \mathfrak{C}_{\mu \nu \rho}$ and $T_{\mu \nu} \ne T_{\nu \mu}$ in general~\cite{NL1, NL2, NL3, NL4, NL5, NL6, NL7, NL8}; however, in this paper, $T_{\mu \nu}$ is symmetric, as in GR, but then it turns out that the \emph{linearized} field equations of NLG are in general inconsistent with $X_{\mu \nu \rho}= \mathfrak{C}_{\mu \nu \rho}$, as demonstrated in Appendix B. To maintain consistency, we therefore assume that $p \ne 0$ in this paper. It will turn out that the tetrad potentials and hence $p$ are only significant for time-varying gravitational fields near their sources. That is, tetrad potentials are negligible for steady-state configurations, see Sec. V. Thus we expect that $p$ can be eventually determined from observational data regarding the gravitational physics of variable sources. 

The constitutive kernel ${\cal K}(x, x')$ could in general depend upon scalars at $x$ and $x'$ that can be formed from the gravitational potentials, the world function $\Omega(x, x')$ and their derivatives. For instance, we can tentatively assume that ${\cal K}(x, x')$ is simply a function of $\Omega_{\mu}(x, x')e^{\mu}{}_{\hat{\alpha}}(x)$ and $\Omega_{\mu'}(x, x')e^{\mu'}{}_{\hat{\alpha}}(x')$, where the Lorentz freedom in the choice of the preferred frame has been fixed relative to the rest frame of the gravitational source as in the following section, where the consequences of this form for  ${\cal K}(x, x')$ are worked out in detail within the framework of the linearized theory. 

It is not known at present whether the field equations of nonlocal gravity can be derived from a variational principle. Moreover, the theory is incomplete without a thorough examination of the physical origin of the nonlocal kernel ${\cal K}$. As discussed in the next section, we take the view that at present the kernel can be determined from observational data regarding dark matter. Perhaps ${\cal K}$ will be ascertained someday from a  more complete future theory. For instance, nonlocality can arise from integrating out certain physical degrees of freedom~\cite{Ga}. 

Nonlocality---in the sense of an influence (``memory") from the past that endures---could be a natural feature of the universal gravitational interaction. Some of the consequences of our nonlocal gravity model have been considered thus far only in the linear weak-field regime~\cite{NL1, NL2, NL3, NL4, NL5, NL6, NL7,NL8}. This has involved detailed studies of the nonlocal modifications of Newtonian gravity and linearized gravitational waves; indeed, these important results are confirmed here via the  approach adopted in the present work. As explained in the following section, the notion that nonlocal gravity simulates dark matter is completely consistent with causality; moreover, the theoretical results appear to be consistent with experiment at the linear level. The nonlinear regime of NLG has not yet been studied; therefore, exact cosmological models or issues involving the influence of nonlocality on the formation and evolution of black holes are beyond the scope of our present considerations.

\section{Linearized Nonlocal Gravity} 

Imagine a finite source of mass-energy in a compact region of space. We suppose that the gravitational field is everywhere weak and falls off to zero far away from the source. We also set $\Lambda=0$ and assume that if gravity is turned off, we are in the rest frame of the source in Minkowski spacetime with the preferred tetrad frame  $e^\mu{}_{\hat{\alpha}}=\delta^\mu_\alpha$. In the presence of gravity, the preferred frame field of nonlocal gravity is then  assumed to be 
\begin{equation}\label{III1}
 e_\mu{}^{\hat{\alpha}}={\d}_\mu ^{\alpha}+\psi^{\alpha}{}_\mu\,, \quad  e^\mu{}_{\hat{\alpha}}=\d^\mu _{\alpha} -\psi^\mu{}_{\alpha}\,,
\end{equation}
where $\psi_{\mu \nu}$ is treated to linear order in perturbation away from Minkowski spacetime and hence the distinction between spacetime and tetrad indices disappears at this level of approximation. Let us note that in Eq.~\eqref{III1}, the invariance of the theory under global Lorentz transformations has been broken, since the preferred frame field coincides with the rest frame of the gravitational source. It is useful to decompose $\psi_{\mu \nu}$  into its symmetric and antisymmetric components; that is, we define, 
\begin{equation}\label{III2}
 h_{\mu \nu}:=2\psi_{(\mu \nu)}, \qquad  \phi_{\mu \nu}:=2\psi_{[\mu \nu]}\,.
\end{equation}
It then follows from Eq.~\eqref{II8} that
\begin{equation}\label{III3} 
g_{\mu \nu}=\eta_{\mu \nu}+h_{\mu \nu}\,.
\end{equation}
Moreover, it is convenient to employ the trace-reversed potentials
\begin{equation}\label{III4}
\overline{h}_{\mu \nu}=h_{\mu \nu}-\frac 12\eta_{\mu \nu}h\,, \qquad   h:=\eta_{\mu \nu}h^{\mu \nu}\,,
\end{equation} 
just as in GR. Here $\overline{h}=-h$ and we have
\begin{equation}\label{III5}
\psi_{\mu \nu}=\frac{1}{2}\overline{h}_{\mu \nu}+\frac 12\phi_{\mu \nu}-\frac 14\eta_{\mu \nu}\overline{h}\,.
\end{equation} 

It is now straightforward to work out the field components in terms of  $\psi_{\mu \nu}$. The torsion tensor is then,  
\begin{equation}\label{III6}
C_{\mu \nu \sigma}=\partial_\mu \psi_{\sigma \nu}-\partial_\nu \psi_{\sigma \mu}
\end{equation}
and the auxiliary torsion tensor is given by 
\begin{equation}\label{III7}
\mathfrak{C}_{\mu \sigma \nu}=-\overline{h}_{\nu [\mu,\sigma]}-\eta_{\nu [\mu}\overline{h}_{\sigma ]\rho,}{}^\rho+\frac 12\phi_{\mu \sigma, \nu}+\eta_{\nu [\mu} \phi_{\sigma ] \rho,}{}^\rho\,,
\end{equation}
in terms of which the Einstein tensor can be expressed as 
\begin{equation}\label{III8}
^{0}G_{\mu \nu}=\partial_\sigma \mathfrak{C}_{\mu}{}^{\sigma}{}_{\nu}=-\frac
12\,\square\, 
\overline{h}_{\mu \nu}+\overline{h}^\rho{}_{(\mu,\nu)\rho}-\frac
12\eta_{\mu \nu}\overline{h}^{\rho \sigma}{}_{,\rho \sigma}\,,
\end{equation}
where $\square :=\eta^{\alpha \beta}\partial_\alpha \partial_\beta$. Moreover, in the linear regime, Eq.~\eqref{II26} reduces to
\begin{equation}\label{III9}
N_{\mu}{}^{\sigma}{}_{\nu}= \int{\cal K}(x, y) X_{\mu}{}^{\sigma}{}_{\nu}(y)~d^4y\,                 
\end{equation}
and $Q_{\mu \nu}$ vanishes. Thus the linearized forms of the field Eqs.~\eqref{II31} and~\eqref{II32} of nonlocal gravity are given by
\begin{equation}\label{III10}
 ^{0}G_{\mu \nu}+\frac{1}{2}\, \partial_\sigma\,(N_{\mu}{}^{\sigma}{}_{\nu}+N_{\nu}{}^{\sigma}{}_{\mu})= \kappa\,  T_{\mu \nu}\,              
\end{equation}
and
\begin{equation}\label{III11}
  \partial_\sigma\,N_{\mu}{}^{\sigma}{}_{\nu}= \partial_\sigma\, N_{\nu}{}^{\sigma}{}_{\mu}\,,              
\end{equation}
respectively. It follows immediately from the antisymmetry of the auxiliary torsion tensor in its first two indices in Eq.~\eqref{III8} and the symmetry of Einstein's tensor that $\partial_{\nu}\, ^{0}G^{\mu \nu}=0$, as expected. Furthermore, Eqs.~\eqref{III10}--\eqref{III11} imply that 
\begin{equation}\label{III11a}
\partial_{\nu} T^{\mu \nu}=0\,,              
\end{equation}
since $N^{\mu \sigma \nu}=-N^{\sigma \mu \nu}$. We thus recover the energy-momentum conservation law for mass-energy, just as in linearized GR. 

Let us next discuss the gauge freedom of the gravitational potentials. An infinitesimal coordinate transformation, $x^\mu \mapsto x'^\mu=x^\mu-\epsilon^\mu(x)$, leads to $\psi_{\mu \nu} \mapsto \psi'_{\mu \nu}=\psi_{\mu \nu}+\epsilon_{\mu,\nu}$ that is valid to linear order in $\epsilon^\mu$. Thus under a gauge transformation,
\begin{equation}\label{III12}
\overline{h}\,'_{\mu \nu}=\overline{h}_{\mu \nu}+\epsilon_{\mu,\nu}+\epsilon_{\nu,\mu}-\eta_{\mu \nu}\epsilon^\alpha{}_{,\alpha}\,, \qquad   \phi'_{\mu \nu}=\phi_{\mu \nu}+\epsilon_{\mu,\nu}-\epsilon_{\nu,\mu}
\end{equation} 
and $\overline{h}\,'=\overline{h}-2\epsilon^\alpha{}_{,\alpha}$; however, as expected, the gravitational field tensors $C_{\mu \nu \sigma}$ and $\mathfrak{C}_{\mu \sigma \nu}$ are left unchanged. It follows that the linearized gravitational field equations of NLG are gauge invariant.
 
To proceed further, we must discuss the nature of the nonlocal kernel in the linearized theory.  The kernel that appears in Eq.~\eqref{III9} is the nonlocal kernel in the Minkowski spacetime limit.  In Minkowski spacetime, the world function is given by~\cite{Sy}
\begin{equation}\label{III13}
\Omega (x, x')=\frac{1}{2} \eta_{\alpha \beta}(x^{\alpha} - x'^{\alpha})(x^{\beta}-x'^{\beta})\,,
\end{equation}
so that to lowest order in the perturbation, we find 
\begin{equation}\label{III14}
\Omega_{\mu}(x, x')e^{\mu}{}_{\hat{\alpha}}(x)=-\Omega_{\mu'}(x, x')e^{\mu'}{}_{\hat{\alpha}}(x')= \eta_{\alpha \beta}(x^{\beta}-x'^{\beta})\,.
\end{equation}
It follows from this result and our brief discussion of the kernel in the previous section that we have a \emph{convolution} kernel in the linearized theory. That is, we can tentatively assume that the nonlocal kernel ${\cal K}(x, y)$ is a \emph{universal} function of $x^{\alpha} - y^{\alpha}$, so that 
\begin{equation}\label{III15}
{\cal K}(x, y):=K(x-y)\,.
\end{equation}
Moreover, to ensure causality, we assume that the convolution kernel $K$ is nonzero only when $x^\alpha-y^\alpha$ is a future directed timelike or null vector in Minkowski spacetime, which means that event $y$ must be within or on the past light cone of event $x$, or equivalently, that event $x$ must be within or on the future light cone of event $y$. That is,   $x^0 \ge y^0$ and 
\begin{equation}\label{III16}
\eta_{\alpha \beta}(x^\alpha-y^\alpha)(x^\beta-y^\beta) \le0\,. 
\end{equation}
It follows that causality is ensured whenever
\begin{equation}\label{III17}
x^0-y^0\ge |\mathbf{x}-\mathbf{y}|\,.
\end{equation} 
Hence, $K(x-y)$ must be proportional to $\Theta(x^0-y^0-|\mathbf{x}-\mathbf{y}|)$, where $\Theta(t)$ is the Heaviside unit step function such that $\Theta(t)=0$ for $t<0$ and $\Theta(t)=1$ for $t\ge 0$. That is,
\begin{equation}\label{III18}
K(x-y) \propto \Theta(x^0-y^0-|\mathbf{x}-\mathbf{y}|)\,.
\end{equation} 

Returning to field Eqs.~\eqref{III10} and~\eqref{III11}, let us now write them more explicitly as follows 
\begin{equation}\label{III18a}
 ^{0}G_{\mu \nu}(x) + \partial_\sigma\,\int K(x-y)\,X_{(\mu}{}^{\sigma}{}_{\nu)}(y)~d^4y= \kappa\,  T_{\mu \nu}(x)\,              
\end{equation}
and
\begin{equation}\label{III18b}
 \partial_\sigma\,\int K(x-y)\, X_{[\mu}{}^{\sigma}{}_{\nu]}(y)~d^4y=0\,.
\end{equation}
The consequences of these equations for various choices of $X_{\mu \sigma \nu}$ are briefly discussed in Appendix B. In this work, however, we choose Eq.~\eqref{II35}, namely, $X_{\mu \sigma \nu}= \mathfrak{C}_{\mu \sigma \nu}+ p\,(\check{C}_\mu\, g_{\sigma \nu}-\check{C}_\sigma\, g_{\mu \nu})$ with $p\ne 0$. Then, in the \emph{linear} regime we have
\begin{equation}\label{III18c}
X_{(\mu}{}^{\sigma}{}_{\nu)} = \mathfrak{C}_{(\mu}{}^{\sigma}{}_{\nu)}+p\,\big[\check{C}_{(\mu}\delta^\sigma_{\nu)}-\check{C}^\sigma \eta_{\mu \nu}\big]\,, \qquad X_{[\mu}{}^{\sigma}{}_{\nu]}= \mathfrak{C}_{[\mu}{}^{\sigma}{}_{\nu]}+p\,\check{C}_{[\mu}\delta^\sigma_{\nu]}\,.
\end{equation}
Let us recall here the fact that the torsion pseudovector $\check{C}^\sigma$ is the dual of $C_{[\mu \nu \rho]}$, which in the linear approximation is given by $C_{[\mu \nu \rho]}=-\phi_{[\mu \nu , \rho]}$. Moreover, in the linear approximation, $\check{C}^{\sigma}{}_{,\sigma}=0$. Thus the part of the constitutive relation proportional to $p$ is given exclusively by the derivatives of tetrad potentials and vanishes for $\phi_{\mu \nu}=0$.

In the calculation of the nonlocal term in Eq.~\eqref{III18a}, $\partial K/\partial x^{\sigma} =-\partial K/\partial y^{\sigma}$, which together with Eq.~\eqref{III8} implies, via integration by parts, that
\begin{equation}\label{III19}
\partial_\sigma\,\int K(x-y)\, \mathfrak{C}_{\mu}{}^{\sigma}{}_{\nu}(y)~d^4y=-S_{\mu \nu} + \int K(x-y)\, ^{0}G_{\mu \nu}(y)~d^4y\,,   
\end{equation}
where $S_{\mu \nu}$ is given by
\begin{equation}\label{III20}
S_{\mu \nu}= \int \frac{\partial}{\partial y^{\sigma}}\Big[K(x-y) \mathfrak{C}_{\mu}{}^{\sigma}{}_{\nu}(y)\Big]~d^4y\,.                 
\end{equation}
Gauss's theorem then implies that 
\begin{equation}\label{III21}
S_{\mu \nu}= \oint K(x-y) \mathfrak{C}_{\mu}{}^{\alpha}{}_{\nu}(y)\,d^3\Sigma_{\alpha}(y)\,,                 
\end{equation}
where the only contribution to the integral comes from the boundary hypersurface at the light cone given by $y^0=x^0-|\mathbf{x}-\mathbf{y}|$.  Therefore, 
\begin{equation}\label{III22}
 S_{\mu \nu}(x)=\int K(|\mathbf{x}-\mathbf{y}|, \mathbf{x}-\mathbf{y})\,\mathfrak{C}_{\mu}{}^0{}_{\nu}(x^0-|\mathbf{x}-\mathbf{y}|, \mathbf{y})~d^3y\,,          
\end{equation}
where $\mathfrak{C}_{\mu}{}^0{}_{\nu}=\mathfrak{C}_{(\mu}{}^0{}_{\nu)}+\mathfrak{C}_{[\mu}{}^0{}_{\nu]}$ is given by Eq.~\eqref{III7}, namely,
\begin{equation}\label{III23}
\mathfrak{C}_{(\mu}{}^{0}{}_{\nu)}=\frac{1}{2}\,\big(\,\overline{h}_{\mu \nu , 0}-\overline{h}_{0 (\mu , \nu)}+ \eta_{\mu \nu}\,\overline{h}_{0 \rho,}{}^{\rho}-\eta_{0 (\mu}\,\overline{h}_{\nu) \rho,}{}^{\rho}+\phi_{0 (\mu , \nu)}-\eta_{\mu \nu}\,\phi_{0 \rho,}{}^{\rho}+\eta_{0 (\mu}\,\phi_{\nu) \rho,}{}^{\rho}\big)\,
\end{equation}
and
\begin{equation}\label{III23a}
\mathfrak{C}_{[\mu}{}^{0}{}_{\nu]}=\frac{1}{2}\,\big(\,\overline{h}_{0[\mu, \nu]}+\phi_{0 [\mu , \nu]}+\eta_{0 [\mu}\,\overline{h}_{\nu] \rho,}{}^{\rho}-\eta_{0 [\mu}\,\phi_{\nu] \rho,}{}^{\rho}\big)\,.
\end{equation}
In a similar way, we find that
\begin{equation}\label{III23b}
U_{\mu \nu} := \partial_\sigma\,\int K(x-y)\,\big(\check{C}_{\mu}\,\delta^\sigma_{\nu}-\check{C}^\sigma\, \eta_{\mu \nu}\big)(y)~d^4y\,
\end{equation}
can be written as 
\begin{equation}\label{III23c}
U_{\mu \nu}= -\int K(|\mathbf{x}-\mathbf{y}|, \mathbf{x}-\mathbf{y})\,\big(\check{C}_{\mu}\,\delta^0_{\nu}-\check{C}^0\, \eta_{\mu \nu}\big)(x^0-|\mathbf{x}-\mathbf{y}|, \mathbf{y})~d^3y+\int K(x-y)\check{C}_{\mu,\nu}(y)~d^4y\,.
\end{equation}
We recall here that $U_{\mu \nu}$ depends only upon the derivatives of  $\phi_{\mu \nu}$ and vanishes for  $\phi_{\mu \nu}=0$.

It follows from these results that in the linear regime, Eq.~\eqref{II29}, which is the nonlocal extension of Einstein's field equations, can be written as
\begin{equation}\label{III24}
 ^{0}G_{\mu \nu}(x)+\int  K(x-y)\, ^{0}G_{\mu \nu}(y)~d^4y= \kappa\,  T_{\mu \nu}(x)+S_{\mu \nu}(x)-p\,U_{\mu \nu}(x)\,.             
\end{equation}
This is the main field equation of linearized nonlocal gravity and can be split into its symmetric and antisymmetric components, namely, 
\begin{equation}\label{III24'}
 ^{0}G_{\mu \nu}(x)+\int  K(x-y)\, ^{0}G_{\mu \nu}(y)~d^4y= \kappa\,  T_{\mu \nu}(x)+S_{(\mu \nu)}(x)-p\,U_{(\mu \nu)}(x)\,             
\end{equation}
and
\begin{equation}\label{III24a}
S_{[\mu \nu]}(x)=p\,U_{[\mu \nu]}(x)\,.
\end{equation}
Let us first note here that $S_{0\nu}(x)=0$ due to the antisymmetry of $\mathfrak{C}_{\mu \sigma \nu}$ in its first two indices. Moreover, it proves useful to define the quantity
\begin{equation}\label{III24aA}
{\mathcal W}_i := -\overline{h}_{00, i}+\overline{h}_{ij,}{}^{j}-\phi_{ij,}{}^{j}\,.
\end{equation} 
Then, the purely nonlocal source-free integral constraints~\eqref{III24a} consist of 6 equations given by
\begin{equation}\label{III24b}
\int K_c(x-y)\, {\mathcal W}_i(y)~d^4y=4\,p\, U_{[i\,0]}(x)\,
\end{equation} 
and 
\begin{equation}\label{III24c}
\int K_c(x-y)\, \big(\,\overline{h}_{0i, j}+\phi_{0i,j}-\overline{h}_{0j,i}-\phi_{0j,i}\big)(y)~d^4y=4\,p\,U_{[i\,j]}(x)\,.
\end{equation} 
Here, we have introduced, for the sake of simplicity, the light-cone kernel $K_c$,
\begin{equation}\label{III24d}
K_c(x-y) :=K(x-y)\,\delta(x^0-y^0-|\mathbf{x}-\mathbf{y}|)\,.         
\end{equation} 
Furthermore, from $S_{0\nu}=0$ and Eq.~\eqref{III24}, we have that 
\begin{equation}\label{III24dA}
 ^{0}G_{0\nu}(x)+\int  K(x-y)\, ^{0}G_{0 \nu}(y)~d^4y= \kappa\,  T_{0\nu}(x)-p\,U_{0\nu}(x)\,,            
\end{equation}
where $U_{0\nu}$ can be determined from Eq.~\eqref{III23c}, namely, 
\begin{equation}\label{III24dB}
U_{0\nu}(x) = \int K(x-y)\check{C}_{0,\nu}(y)~d^4y\,.
\end{equation}
In Appendix B, we show that $\check{C}_0$ can be determined in principle in terms of $T_{00}$, see Eq.~\eqref{B30}. Finally, the source term for the field equation involving $^{0}G_{ij}$ contains $S_{(ij)}$ and $U_{(ij)}$, where
\begin{equation}\label{III24e}
S_{(ij)}(x)=\frac{1}{2}\,\int K_c(x-y)\, \big[\,\overline{h}_{ij, 0}-\overline{h}_{0(i,j)}+\phi_{0(i,j)}+\delta_{ij}\,(\overline{h}_{0\rho,}{}^{\rho}-\phi_{0k,}{}^{k})\big](y)~d^4y\,         
\end{equation} 
and $U_{(ij)}$ can be simply determined from Eq.~\eqref{III23c}.

It is clear from these results that in our decomposition of the linear gravitational potentials $\psi_{\mu \nu}$ in Eq.~\eqref{III2},  the symmetric \emph{metric} part $\overline{h}_{\mu \nu}$ that satisfies Eq.~\eqref{III24'} has primary dynamical content, while the antisymmetric \emph{tetrad} part $\phi_{\mu \nu}$ plays a secondary role and is constrained via  Eq.~\eqref{III24a}. In general, $\overline{h}_{\mu \nu}$ and $\phi_{\mu \nu}$ are \emph{inextricably connected} in both sets of equations and cannot be simply disentangled. In the case of $X_{\mu \nu \rho}= \mathfrak{C}_{\mu \nu \rho}+ p\,(\check{C}_\mu\, g_{\nu \rho}-\check{C}_\nu\, g_{\mu \rho})$ under consideration here, certain simplifications occur that are discussed in the last part of this section. 

Nonlocal gravity has a characteristic galactic  length scale of order 1 kpc; therefore, in the vicinity of a planet, a star or a binary star system, whose dimensions are very small compared to 1 kpc, the nonlocal terms in Eqs.~\eqref{III24'} and~\eqref{III24a} can be generally neglected and linearized nonlocal gravity simply reduces to linearized GR.  Therefore, in the discussion of gravitational radiation of reduced wavelength $\lambdabar \ll 1$ kpc, which is the regime of current observational interest, nonlocal effects in the generation and detection of such waves are essentially negligible~\cite{NL6}. Nonlocal effects can, however, be significant in the galactic or extragalactic propagation of waves from the source to the detector~\cite{NL6, NL7}.  

Before discussing the solution of the linearized field equations, we must digress here and point out a significant consequence of gravitational dynamics given by Eq.~\eqref{III24}. Working in the space of continuous functions on spacetime that are absolutely integrable ($L^1$) as well as square integrable ($L^2$), it is possible to write Eq.~\eqref{III24} in the form
\begin{equation}\label{III25}
 ^{0}G_{\mu \nu}= \kappa\,  T_{\mu \nu}+S_{\mu \nu}-p\,U_{\mu \nu}+\int  R(x-y)\, [\kappa \, T_{\mu \nu}+S_{\mu \nu}-p\,U_{\mu \nu}](y)~d^4y\,,             
\end{equation}
where $R(x-y)$ is a kernel that is \emph{reciprocal} to $K(x-y)$~\cite{Tr}. The reciprocal kernel is of the \emph{convolution} type and is \emph{causal} as well. Aside from nonlocal terms involving $S_{\mu \nu}$ and $U_{\mu \nu}$, Eq.~\eqref{III25} exhibits an important feature that must be stressed. That the linearized gravitational field equations can be expressed as in Eq.~\eqref{III25} is a crucial result, since it means that nonlocal gravity in the linear regime is essentially equivalent to general relativity, except that in addition to the usual gravitational source, there is an additional ``dark" source that is given by the convolution of the usual source with the \emph{causal} reciprocal kernel. In nonlocal gravity theory, this additional source is identified as the main component of what appears as \emph{dark matter} in astrophysics. Thus nonlocality simulates dark matter in this linearized theory, since the latter is simply the manifestation of the nonlocal aspect of the gravitational interaction.

\subsection{Causal Reciprocal Kernel}

Due to the importance of Eq.~\eqref{III25} for the physical interpretation of NLG, this subsection is devoted to a brief description of the mathematical steps that lead to this result. It turns out that the convolution property of the kernels under consideration is independent of their crucial causality properties. Therefore, we first consider a kernel $K(x, y)$ that is causal, so that $K(x, y)$ vanishes unless Eq.~\eqref{III17} is satisfied in this case. 

A \emph{Volterra kernel} is defined to be a \emph{causal} kernel function $K(x, y)$ that is continuous over causally ordered sets in Minkowski spacetime.  The product of two Volterra kernels $K$ and $K'$ is defined  to be 
\begin{equation}\label{III26}
V(x,y) =  \int_{{\cal D}(x,y)} K(x,z) K'(z,y)~d^4z\,, 
\end{equation}
which is a Volterra kernel, since the above integrand is nonzero only when $z$ is simultaneously in the past light cone of $x$ and in the future light cone of $y$, so that $y$ is in the past light cone of $x$. Thus the integration domain ${\cal D}(x, y)$ in Eq.~\eqref{III26} is the \emph{finite} region in Minkowski spacetime bounded by the past light cone of event $x$ and the future light cone of event $y$. Alternatively, consider the causality conditions for $K$ and $K'$, namely, 
\begin{equation}\label{III27}
x^0-z^0\ge |\mathbf{x}-\mathbf{z}|\,,\qquad  z^0-y^0\ge |\mathbf{z}-\mathbf{y}|\,,
\end{equation}
respectively. These imply, via addition, that $V$ is causal, since
\begin{equation}\label{III28}
x^0-y^0\ge |\mathbf{x}-\mathbf{z}|+ |\mathbf{z}-\mathbf{y}|\ge |\mathbf{x}-\mathbf{y}|\,,
\end{equation}
by the triangle inequality. Volterra kernels thus form an \emph{algebra} over the causally ordered events in Minkowski spacetime. 

Consider next the generalized Volterra integral equation of the second kind given by
\begin{equation}\label{III29}
 B(x,y)+ \int_{{\cal D}(x,y)} K(x,z)\, B(z,y)~d^4z =  A(x,y)\,,
\end{equation}
where $A(x,y)$ and $K(x,y)$ are given Volterra kernels and we wish to find a Volterra kernel $B(x,y)$ that satisfies this equation. According to a general theorem due to M. Riesz~\cite{MR, FV}, there is 
 a \emph{unique solution} given by
\begin{equation}\label{III30}
 A(x,y)+ \int_{{\cal D}(x,y)} R(x,z)\,A(z,y)~d^4z = B(x,y)\,,
\end{equation}
where the reciprocal Volterra kernel $R(x,y)$ can be expressed as
\begin{equation}\label{III31}
R(x,y) = \sum_{n=1}^{\infty} K_n(x,y)\,.
\end{equation}
Here the iterated Volterra kernels $K_n(x,y)$ for $n=1,2,3,...$ are defined such that $K_1(x,y):=-K(x,y)$ and 
\begin{equation}\label{III32}
K_{n+1}(x,y) :=  \int_{{\cal D}(x,y)} K_n(x,z)\, K_1(z,y)~d^4z\,. 
\end{equation}
The Neumann series~\eqref{III31} converges uniformly on bounded domains and the reciprocal kernel $R$ is indeed a Volterra kernel.  This is proved in the paper of  Faraut and Viano~\cite{FV} using generalized Riemann-Liouville kernels. The work of M. Riesz~\cite{MR} employed a wider context; here, we have followed the treatment of Ref.~\cite{FV}.

It is simple to demonstrate that this significant mathematical result holds just as well if Volterra kernels are all of the convolution type; that is, we can replace $K(x,y)$ by $K(x-y)$, etc. For instance, a simple change of variable in the corresponding integral in Eq.~\eqref{III26} is enough to show that $V$, the product of Volterra kernels $K$ and $K'$ of convolution type, is also of convolution type and that, furthermore,  $V$ is also the product of $K'$ and $K$.  Therefore, \emph{convolution Volterra kernels} form a \emph{commutative subalgebra} of the Volterra algebra. 

Henceforth, we limit our considerations  to Volterra \emph{convolution} kernels that are $L^1$ and $L^2$ functions on spacetime. We wish to reduce the generalized Volterra integral Eqs.~\eqref{III29} and~\eqref{III30} to the following Volterra integral equations: 
\begin{equation}\label{III33}
{\cal G}(x)+ \int K(x-y)\, {\cal G}(y)~d^4y = {\cal F}(x)\,
\end{equation}
and
\begin{equation}\label{III34}
{\cal F}(x)+ \int R(x-y)\, {\cal F}(y)~d^4y = {\cal G}(x)\,.
\end{equation}
To this end, consider  any continuous $L^1$ function  $f(x)$ over spacetime and define 
\begin{equation}\label{III35}
{\cal F}(x) :=\int A(x-y) f(y)~d^4y\,, \qquad  {\cal G}(x) := \int B(x-y) f(y)~d^4y\,,
\end{equation}
where $A$ and $B$ are closely related to the Volterra kernels defined in Eqs.~\eqref{III29} and~\eqref{III30}. That is, replacing the kernels in Eqs.~\eqref{III29} and~\eqref{III30} by $L^1$ and $L^2$ \emph{convolution} kernels, multiplying the resulting equations by $f(y)$ and integrating over spacetime, we obtain Eqs.~\eqref{III33} and~\eqref{III34}. It is a simple consequence of Young's inequality for convolutions that if $f$ and $A$ are $L^1$ functions, then their convolution ${\cal F}$ is also $L^1$. Thus we find that in Eq.~\eqref{III35}, ${\cal F}(x)$ and ${\cal G}(x)$ are continuous $L^1$ functions over spacetime. Moreover, it follows from Minkowski's integral inequality that if $f$ is $L^1$ and $A$ is $L^2$, then their convolution is $L^2$. Hence, ${\cal F}(x)$ and ${\cal G}(x)$ are $L^2$ functions over spacetime as well. 

The substitution of Eq.~\eqref{III33} into Eq.~\eqref{III34}, or vice versa, results in the basic \emph{reciprocity} integral equation
\begin{equation}\label{III35a}
K(x-y)+R(x-y)+ \int K(x-z)R(z-y)~d^4z = 0\,.
\end{equation}
It is clear that the convolution Volterra kernels $K$ and $R$ can be interchanged in this reciprocity relation. 

Writing  ${\cal G}$ for $^{0}G_{\mu \nu}$  and ${\cal F}$ for $\kappa\,  T_{\mu \nu}+S_{\mu \nu}-p\,U_{\mu \nu}$ in Eq.~\eqref{III24}, we recover Eq.~\eqref{III33}, which means that Eq.~\eqref{III34} is then equivalent to Eq.~\eqref{III25}; in particular, we have the remarkable result that in the space of continuous and absolutely integrable as well as square integrable functions on spacetime, the reciprocal kernel exists and is causal, so that 
\begin{equation}\label{III36}
R(x-y) \propto \Theta(x^0-y^0- |\mathbf{x}-\mathbf{y}|)\,.
\end{equation}
Furthermore, it is possible to express Eqs.~\eqref{III33} and~\eqref{III34} in the Fourier domain. That is, let
\begin{equation}\label{III37}
\hat{f} (\xi) =  \int f(x) e^{-i \xi \cdot x}~ d^4x\,
\end{equation}
be the Fourier transform of $f$ in spacetime, where $\xi \cdot x := \eta_{\alpha \beta}\xi^\alpha x^\beta$. Then, 
\begin{equation}\label{III38}
f(x) = \frac{1}{(2\pi)^4} \int \hat{f} (\xi) e^{i \xi \cdot x}~ d^4\xi\,.
\end{equation}
It follows from the convolution theorem for Fourier transforms that Eqs.~\eqref{III33} and~\eqref{III34} can be written in the Fourier domain as $\hat{{\cal F}}= \hat{{\cal G}}(1+\hat{K})$ and $\hat{{\cal G}}= \hat{{\cal F}}(1+\hat{R})$, respectively. Therefore, 
\begin{equation}\label{III39}
(1+\hat{K})(1+\hat{R})=1\,,
\end{equation}
which can also be obtained directly via Fourier transformation from Eq.~\eqref{III35a} and is an expression of the complete reciprocity between $K$ and $R$. In particular, suppose that $R(x-y)$ can be estimated from the observational data regarding dark matter, then the kernel of nonlocal gravity $K(x-y)$ can be determined from the Fourier transform of 
\begin{equation}\label{III40}
\hat{K} (\xi) =- \frac{\hat{R} (\xi)}{1+\hat{R} (\xi)}\,, 
\end{equation}
 provided $1+\hat{R} (\xi)\ne 0$.
 
 \subsection{Linearized Field Equations with $\overline{h}^{\mu\nu}{}_{, \nu}=0$}
 
Let us now return to Eqs.~\eqref{III24}--\eqref{III25} that characterize linearized nonlocal gravity and  discuss the general structure and the formal solution of the nonlocal field equations for the gravitational field of an isolated source.  For $K=R=0$ in these equations, nonlocality disappears and the field equations reduce to the familiar second-order partial differential equations of linearized GR.  We assume, for the present discussion, that kernels $K$ and $R$ are known; in fact, their determination is the subject of the next section. 

In connection with Eq.~\eqref{III25}, it is useful to define the \emph{total} matter energy-momentum tensor ${\cal T}_{\mu \nu}$,
\begin{equation}\label{III41}
{\cal T}_{\mu \nu} :=T_{\mu \nu}+T^{D}_{\mu \nu}\,,
\end{equation} 
where $T^{D}_{\mu \nu}$, the convolution of  $T_{\mu \nu}$ and $R$, is the ``dark" counterpart of the matter energy-momentum tensor $T_{\mu \nu}$. That is, 
\begin{equation}\label{III42}
T^{D}_{\mu \nu}(x)= \int R(x-y)\,T_{\mu \nu}(y)~d^4y\,.
\end{equation} 
Similarly, we define
\begin{equation}\label{III43}
{\cal S}_{\mu \nu}(x):= S_{\mu \nu}(x)+\int R(x-y)\,S_{\mu \nu}(y)~d^4y\,
\end{equation}
and
\begin{equation}\label{III43a}
{\cal U}_{\mu \nu}(x):= U_{\mu \nu}(x)+\int R(x-y)\,U_{\mu \nu}(y)~d^4y\,.
\end{equation}  
It is possible to write these equations as
\begin{equation}\label{III43b}
{\cal S}_{\mu \nu}(x)= \int W(x-y)\,\mathfrak{C}_{\mu}{}^{0}{}_{\nu}(y)~d^4y\,,
\end{equation}
where $\mathfrak{C}_{\mu}{}^{0}{}_{\nu}$ is given by Eqs.~\eqref{III23}--\eqref{III23a}, and
\begin{equation}\label{III43c}
{\cal U}_{\mu \nu}(x)= -\int W(x-y)\,\big(\check{C}_{\mu}\,\delta^0_{\nu}-\check{C}^0\, \eta_{\mu \nu}\big)(y)~d^4y-\int R(x-y)\,\check{C}_{\mu,\nu}(y)~d^4y\,.
\end{equation}  
Here, we have introduced convolution kernel W,
\begin{equation}\label{III43d}
W(x-y) :=K_c(x-y)\,+\int R(x-z)K_c(z-y)~d^4z\,,        
\end{equation} 
where  in the integrand $R$ and $K_c$ can be interchanged. Moreover, in deriving Eq.~\eqref{III43c}, we have used the reciprocity relation~\eqref{III35a}.

As in GR, the gauge freedom of the gravitational potentials may be used to impose the transverse gauge condition  
\begin{equation}\label{III44}
\overline{h}^{\mu\nu}{}_{, \nu}=0\,. 
\end{equation} 
The remaining gauge degrees of freedom involve four functions $\epsilon^\mu(x)$ such that  $\Box \epsilon^\mu=0$. 
With the imposition of the transverse gauge condition, we find from Eq.~\eqref{III8} that 
\begin{equation}\label{III45}
^{0}G_{\mu \nu}=-\frac{1}{2}  \Box \overline{h}_{\mu\nu}\,.
\end{equation} 
Hence, our main dynamical result, Eq.~\eqref{III25}, can be expressed as
\begin{equation}\label{III45a}
 \Box \overline{h}_{\mu \nu}+ 2{\cal S}_{\mu \nu} =-2\kappa \, {\cal T}_{\mu \nu}+ 2 p\,{\cal U}_{\mu \nu}\,.
\end{equation} 
That is, 
\begin{equation}\label{III46}
 \Box \overline{h}_{0\mu}=-2\kappa \, {\cal T}_{0\mu}-2p\,\int R(x-y)\,\check{C}_{0,\mu}(y)~d^4y\,,
\end{equation} 
since $S_{0\mu}=0$ and hence ${\mathcal S}_{0\mu}=0$ as well. Furthermore, 
\begin{equation}\label{III47}
 \Box \overline{h}_{ij} + \int W(x-y)\big[\,\overline{h}_{ij, 0}-\overline{h}_{0(i,j)}+\phi_{0(i,j)}-\delta_{ij}\,\phi_{0k,}{}^{k}\big](y)~d^4y=-2\kappa \, {\cal T}_{ij}+2p\,{\cal U}_{(ij)}\,,
\end{equation} 
where
\begin{equation}\label{III48}
{\cal U}_{(ij)}(x)= -\delta_{ij}\int W(x-y)\,\check{C}_0(y)~d^4y-\int R(x-y)\,\check{C}_{(i,j)}(y)~d^4y\,.
\end{equation}  
We must solve these dynamic field equations subject to the 6 integral constraints given by Eqs.~\eqref{III24b} and~\eqref{III24c}. Once the 10 components of $\overline{h}_{\mu \nu}$ have been determined, one can find the metric perturbation
\begin{equation}\label{III49}
h_{\mu \nu}= \overline{h}_{\mu \nu}-\frac{1}{2} \eta_{\mu \nu} \overline{h}\,.
\end{equation} 
On the other hand, the constraints appear to be dominated by $\phi_{\mu \nu}=-\phi_{\nu \mu}$. Let us recall that the gravitational potentials of linearized nonlocal gravity, $\psi_{\mu \nu}=\psi_{(\mu \nu)}+\psi_{[\mu \nu]}$, consist of 10 metric variables $\psi_{(\mu \nu)}=\frac{1}{2}h_{\mu \nu}$ and 6 tetrad variables $\psi_{[\mu \nu]}=\frac{1}{2}\phi_{\mu \nu}$. These variables are all intertwined in the linearized field equations of NLG.

It is shown in Appendix B that the field equation for $ \overline{h}_{00}$ can be combined with constraint~\eqref{III24b} to derive Eq.~\eqref{B30} for $\check{C}_0 = O(c^{-2})$.  Assuming that $\check{C}_0$ can be determined in terms of $T_{00}$ from Eq.~\eqref{B30}, we can then calculate ${\mathcal U}_{0\mu}$  via
\begin{equation}\label{III50a}
{\mathcal U}_{0\mu}=-\int R(x-y) \check{C}_{0,\mu}(y)~d^4y\,.
\end{equation} 
The general solution of Eq.~\eqref{III46} involves the superposition of a particular solution of the inhomogeneous equation plus a general solution of the wave equation.  Assuming the absence of incoming gravitational waves, we are interested in the special retarded solution
\begin{equation}\label{III50b} 
\overline{h}_{0\mu}(x^0, \mathbf{x})=\frac{\kappa}{2 \pi}\int \frac{\big[\,{\cal T}_{0\mu}-(p/\kappa)\,{\mathcal U}_{0\mu}\,\big](x^0-|\mathbf{x}-\mathbf{y}|, \mathbf {y})}{|\mathbf{x}-\mathbf{y}|}~d^3y\,.
\end{equation}
The other variables cannot be simply decoupled in general.  

In connection with the propagation of gravitational waves, let us note that very far from the source, where ${\mathcal T}_{\mu \nu} \approx 0$, Eqs.~\eqref{III46}--\eqref{III48} and 
constraints~\eqref{III24b}--\eqref{III24c} are consistent in the transverse-traceless (TT) gauge with $ \overline{h}_{0\mu}=0$ and $\phi_{\mu \nu}=0$. Then,
\begin{equation}\label{III50c}
 \Box \overline{h}_{ij} + \int W(x-y)\,\overline{h}_{ij, 0}(y)~d^4y \approx 0\,,
\end{equation} 
in general agreement with  Refs.~\cite{NL6, NL7}.
In this field equation for $\overline{h}_{ij}$, it is interesting to note a \emph{nonlocal damping} feature that has been studied in Ref.~\cite{NL7}. Thinking about Eq.~\eqref{III50c} in terms of a simple analogy with the mechanics of a linear damped oscillator, we note that the term  $\partial \overline{h}_{ij}/\partial t$ in Eq.~\eqref{III50c} is reminiscent of the ``velocity" of the corresponding oscillator. It is interesting that such a nonlocal damping is completely absent in Eq.~\eqref{III46}, which for $\overline{h}_{00}$ is the physical basis for the modified Poisson equation in the Newtonian regime of nonlocal gravity.  The general solution of the linearized field equations of NLG is beyond the scope of this investigation. However, some special cases of particular physical interest are treated in sections V and VI.

To go further, it is necessary to have knowledge of the reciprocal nonlocal kernels $K$ and $R$. This is the subject of the next section.

\section{Reciprocal Kernel $R$ of Linearized NLG}

The reciprocity between the nonlocal kernels $K$ and $R$ implies that it is in principle sufficient to determine only one of them. This section is therefore primarily devoted to the determination of $R$, since it is more directly connected to astrophysical applications. The first step involves the Newtonian limit of nonlocal gravity, which can be used to determine $R$ in the Newtonian regime from the comparison of the theory with observational data regarding dark matter in spiral galaxies as well as clusters of galaxies~\cite{NL8}. 

\subsection{Newtonian Limit} 

The Newtonian regime is marked by instantaneous action at a distance; therefore, it is natural to assume that for $c \to \infty$, gravitational memory is purely spatial and all retardation effects vanish. It follows that in the Newtonian limit
\begin{eqnarray}\label{IV1}
K(x-y)=\delta(x^0-y^0)\,\chi(\mathbf{x}-\mathbf{y})\,
\end{eqnarray}
and then reciprocity requires that
\begin{eqnarray}\label{IV2}
R(x-y)=\delta(x^0-y^0)\,q(\mathbf{x}-\mathbf{y})\,.
\end{eqnarray}
In fact, the substitution of these Newtonian kernels in our basic relations~\eqref{III33} and~\eqref{III34} results in the reciprocity relation for  spatial kernels, namely,
\begin{eqnarray}\label{IV2a}
\chi(\mathbf{x}-\mathbf{y})+q(\mathbf{x}-\mathbf{y})+\int \chi(\mathbf{x}-\mathbf{z})\,q(\mathbf{z}-\mathbf{y})~d^3z=0\,.
\end{eqnarray}
We will assume that these spatial kernels are symmetric in the sense that $\chi(\mathbf{x}-\mathbf{y})$ is only a function of $|\mathbf{x}-\mathbf{y}|$, etc. Thus in the Fourier domain, we have
\begin{eqnarray}\label{IV2b}
{\hat \chi}(|\boldsymbol{\xi}|)+{\hat q}(|\boldsymbol{\xi}|)+{\hat \chi}(|\boldsymbol{\xi}|)\,{\hat q}(|\boldsymbol{\xi}|)=0\,.
\end{eqnarray}

Let us now use Eq.~\eqref{IV2} in the linearized field Eq.~\eqref{III46} to determine the generalization of Poisson's equation of Newtonian gravity as we formally let $c \to \infty$. We assume that the dominant term of the matter energy-momentum tensor is given by $T_{00}=\rho c^2$, where $\rho$ is the density of matter, and $\overline{h}_{0 0}=-4\Phi/c^2$. Moreover, it follows from Eq.~\eqref{B30} of Appendix B that $\check{C}=O(c^{-2})$. Thus,  we find from Eq.~\eqref{III46} that as $c \to \infty$,
\begin{equation}\label{IV3}
\nabla^2\Phi (\mathbf{x}) = 4\pi G(\rho+\rho_D)\,, \qquad \rho_D(\mathbf{x})=\int q(\mathbf{x}-\mathbf{y}) \rho(\mathbf{y})d^3y\,,
\end{equation}
where $\rho_D$ is the density of ``dark" matter and we have suppressed the dependence of $\Phi$, $\rho$ and $\rho_D$ upon time $t$ for the sake of simplicity.  We take the view that dark matter is essentially a consequence of the nonlocal aspect of the gravitational interaction~\cite{NL1, NL2,NL3,NL4,NL5,NL6,NL7,NL8}. That is, nonlocality simulates dark matter at least at the linear order, and hence this nonlocality should be able to account for the observational aspects of the astrophysical phenomena attributed to dark matter. A beginning has already been made in this direction in Ref.~\cite{NL8}, which also contains an essentially complete description of the Newtonian regime of nonlocal gravity. We therefore briefly review here the steps by which $q(\mathbf{x}-\mathbf{y})$ has been determined thus far. 

Starting from the Newtonian laws of motion and taking into account the observational data regarding the nearly flat rotation curves of spiral galaxies~\cite{RF,RW,SR}, one finds that in the absence of dark matter, the Newtonian attraction of gravity on the galactic scale must vary essentially as the inverse of the distance from the center of the galaxy. That is, the gravitational force acting on a star of mass $m$ circling the bulge in the galactic disk  would be essentially $mv_0^2/r$, where $v_0$ is the constant rotation speed that corresponds to the flat rotation curve of the spiral galaxy.  This means that the Newtonian inverse-square law of gravity, which is valid on solar-system scales, must be suitably modified on galactic scales and beyond. Moreover, the spatial kernels $q$ and $\chi$ must be smooth functions of the kind discussed in the previous section. This problem has been dealt with in depth in Ref.~\cite{NL5}, where two simple possible solutions to the problem were investigated in detail. These are 
\begin{equation}\label{IV4}
q_1=\frac{1}{4\pi \lambda_0}~ \frac{1+\mu (a_0+r)}{(a_0+r)^2}~e^{-\mu r}\,
\end{equation}
and
\begin{equation}\label{IV5}
q_2=\frac{1}{4\pi \lambda_0}~ \frac{1+\mu (a_0+r)}{r(a_0+r)}~e^{-\mu r}\,,
\end{equation}
where $r=|\mathbf{x}-\mathbf{y}|$ and $\lambda_0$, $a_0$ and $\mu$ are constant parameters such that $\lambda_0$, the fundamental length scale of NLG, is expected to be of the order of 1~kpc and
\begin{equation} \label{IV6}
0<\mu \lambda_0<1\,, \qquad  0<\mu a_0 \ll1\,, \qquad 0< a_0/\lambda_0 \ll 1\,.
\end{equation}

It turns out that Eqs.~\eqref{IV3}--\eqref{IV5} constitute a generalization of the phenomenological Tohline-Kuhn approach to modified gravity~\cite{T,K,B}; in fact, kernels~\eqref{IV4} and~\eqref{IV5} are suitable generalizations of the Kuhn kernel~\cite{B} within the framework of nonlocal gravity.

In conformity with the requirements of the previous section (cf. subsection A of Sec. III), kernels $q_1$ and $q_2$ are continuous positive functions that are integrable as well as square integrable over all space. The Fourier transform of $q_1$ is a real positive function if $a_0/\lambda_0$ is sufficiently small compared to unity. On the other hand, the Fourier transform of $q_2$ is always real and positive regardless of the value of $a_0/\lambda_0$. These results imply, via the Fourier transform method, that the corresponding kernels $\chi_1$ and $\chi_2$ exist, are symmetric and have other desirable physical properties~\cite{NL5}.

In many situations of physical interest, $a_0/\lambda_0$, $0< a_0/\lambda_0 \ll 1$, can be neglected, in which case $q_1$ and $q_2$ both reduce to~\cite{NL8}
\begin{equation}\label{IV7}
 q_0=\frac{1}{4 \pi \lambda_0}\frac{(1+\mu r)}{r^2}e^{-\mu r}\,,
\end{equation}
which is integrable over all space such that 
\begin{equation}\label{IV8}
\int q_0(\mathbf{x})~d^3x= \alpha\,, \qquad \alpha := \frac{2}{\lambda_0 \mu}\,.
\end{equation}
It is then straightforward to work out, using Eqs.~\eqref{IV3} and~\eqref{IV7}, the nonlocal generalization of Newton's inverse-square law of gravity, namely, 
\begin{equation}\label{IV9}
\mathbf{F}_{NLG}=-\frac{Gm_{\mathbf{x}}m_{\mathbf{y}}\,(\mathbf{x}-\mathbf{y})}{r^3} \Big[1+\alpha-\alpha (1+\frac{1}{2}\mu r)e^{-\mu r}\Big]\,.
\end{equation}
This represents the attractive central conservative force acting on point mass $m_{\mathbf{x}}$ at $\mathbf{x}$ due to the presence of point mass $m_{\mathbf{y}}$ at $\mathbf{y}$. It is interesting to note that $\mathbf{F}_{NLG}$ is a linear superposition of an attractive Newtonian force of gravity augmented by $(1+\alpha)$, where $\alpha \approx 11$, and a repulsive  Yukawa-type force with a spatial galactic decay length of $\mu^{-1} \approx 17$ kpc~\cite{NL8}. Newton's inverse-square force law is recovered when $r$ can be neglected in comparison with $\mu^{-1}$. On the other hand,  on the scales of clusters of galaxies and beyond, where $\mu r \gg 1$, the Yukawa-type force can  be neglected and  the force of gravity is then essentially Newtonian but with $G \to G(1+\alpha)$. Moreover, regarding the \emph{exterior} gravitational field of an extended source, we find from the integration of Eq.~\eqref{IV9} over a spherical mass distribution of radius ${\mathcal R}_0$ that  the mass distribution can, in effect, be treated approximately as a point mass if ${\mathcal R}_0$ is completely negligible compared to $\mu^{-1}$~\cite{NL3, NL8}. 

A detailed investigation has revealed that Eq.~\eqref{IV9} is consistent with gravitational dynamics  in the Solar System, spiral galaxies and clusters of galaxies with 
\begin{equation}\label{IV10}
\alpha = 10.94\pm2.56\,, \qquad  \mu = 0.059\pm0.028~{\rm kpc^{-1}}\,
\end{equation}
and $\lambda_0 \approx 3 \pm 2~ {\rm kpc}$, where $\lambda_0=2/(\alpha \mu)$~\cite{NL8}.

\subsection{Beyond the Newtonian Regime}

Memory generally dies out; therefore, we expect nonlocal kernels $K$ and $R$ to decay exponentially in space and time. The exponential decay term in $q$ already indicates that the distance scale associated with spatial gravitational  memory is $\mu^{-1} \approx 17$ kpc. We should therefore expect a similar temporal behavior in $K$ and $R$; moreover, causality requires that these kernels be proportional to the Heaviside unit step function as in Eqs.~\eqref{III18} and~\eqref{III36}. Thus the Dirac delta function  $\delta(x^0-y^0)$ that appears in Eqs.~\eqref{IV1} and~\eqref{IV2} should be suitably generalized for \emph{finite} $c$ to satisfy these requirements. 

Consider the set of functions $\delta_n(s)$ for $n=1,2,3,...$ given by
\begin{equation}\label{IV11}
\delta_n(s):= \nu ~ n~ e^{-\nu \,n\,(s-\frac{r}{n})}~\Theta\big(s-\frac{r}{n}\big)\,,
\end{equation}
where $\nu>0$ and $r\ge 0$ are constants. These functions are normalized,
\begin{equation}\label{IV12}
\int_{- \infty}^{\infty} \delta_n(s)~ds = 1\,, 
\end{equation}
and form a \emph{Dirac sequence}, since it can be shown that for any smooth function $f(s)$,
\begin{equation}\label{IV13}
\lim_{n \to \infty} \int_{- \infty}^{\infty} \delta_n(s)~f(s)~ds = f(0^+)\,.
\end{equation}
Therefore, the Dirac delta function $\delta(s)$ may be regarded as a certain distributional limit of the sequence of normalized functions $\delta_n(s)$ as $n \to \infty$. Moreover, we note that the singularity of this Dirac delta function occurs at $0^+$, the positive side of the origin. 

In Eq.~\eqref{IV11}, let us now formally replace $s$ by $t_{\mathbf{x}}-t_{\mathbf{y}}$, $r$ by $|\mathbf{x}-\mathbf{y}|$ and $n$ by the speed of light $c$; then, it is straightforward to check that in the limit as $c \to \infty$, we have
\begin{equation}\label{IV14}
 \nu ~ c~ e^{-\nu \,c\,\big(t_{\mathbf{x}}-t_{\mathbf{y}}-\frac{|\mathbf{x}-\mathbf{y}|}{c}\big)}~\Theta\Big(t_{\mathbf{x}}-t_{\mathbf{y}}-\frac{|\mathbf{x}-\mathbf{y}|}{c}\Big)~ \to ~\delta(t_{\mathbf{x}}-t_{\mathbf{y}})\,
\end{equation}
in the distributional sense of Eq.~\eqref{IV13}. It follows from these considerations that when the finite magnitude of the speed of light is taken into account, $\delta(x^0-y^0)$ in Eq.~\eqref{IV2} can be replaced by 
\begin{equation}\label{IV15}
 \nu ~ e^{-\nu\,(x^0-y^0-|\mathbf{x}-\mathbf{y}|)}~\Theta \big(x^0-y^0-|\mathbf{x}-\mathbf{y}|\big)\,,
\end{equation}
where we recall that $x^0=c~ t_{\mathbf{x}}$, $y^0=c~ t_{\mathbf{y}}$ and $\delta(t_{\mathbf{x}}-t_{\mathbf{y}})= c ~\delta(x^0-y^0)$. Here, $\nu^{-1}$ is a constant length that should ultimately be determined on the basis of observational data. As in Ref.~\cite{NL6}, we speculate that $\nu^{-1}$ is a galactic length that is equal to, or comparable with, $\mu^{-1}$.

Henceforward, we assume that 
\begin{equation}\label{IV16}
R(x-y)= \nu ~ e^{-\nu\,(x^0-y^0-|\mathbf{x}-\mathbf{y}|)}~\Theta \big(x^0-y^0-|\mathbf{x}-\mathbf{y}|\big)~q(\mathbf{x}-\mathbf{y})\,.
\end{equation}
This reciprocal kernel $R$ is consistent with our physical requirements and depends only upon $x^0-y^0$ and $|\mathbf{x}-\mathbf{y}|$. An important consequence of the normalization property of Eq.~\eqref{IV15}, namely, 
\begin{equation}\label{IV17}
\int ~ \nu ~ e^{-\nu\,(x^0-y^0-|\mathbf{x}-\mathbf{y}|)}~\Theta \big(x^0-y^0-|\mathbf{x}-\mathbf{y}|\big)~dy^0=1\,,
\end{equation}
is that
\begin{equation}\label{IV17a}
 \int R(x-y)\,Z(\mathbf{y})~d^4y= \int q(\mathbf{x}-\mathbf{y})\,Z(\mathbf{y})~d^3y\,
\end{equation} 
for any smooth purely \emph{spatial} function $Z(\mathbf{x})$. In the Fourier domain, this relation amounts to 
\begin{equation}\label{IV17b}
{\hat R}(0, \boldsymbol{\xi})= {\hat q}(\boldsymbol{\xi})\,,
\end{equation} 
which implies, via Eq.~\eqref{III39}, when $\xi^0=0$, and Eq.~\eqref{IV2b}, that 
\begin{equation}\label{IV17c}
{\hat K}(0, \boldsymbol{\xi})= {\hat \chi}(\boldsymbol{\xi})\,,
\end{equation} 
or, in the spacetime domain,
\begin{equation}\label{IV17d}
 \int K(x-y)\,Z(\mathbf{y})~d^4y= \int \chi(\mathbf{x}-\mathbf{y})\,Z(\mathbf{y})~d^3y\,.
\end{equation} 

Finally, it is interesting to note that for $Z=1$,
the integral of the reciprocal kernel $R$ over the whole spacetime is given by
\begin{equation}\label{IV18}
 \int R(x)~d^4x= \int q(\mathbf{x})~d^3x={\hat q}(0)\,,
\end{equation}
which can be easily computed for $q_1$ and $q_2$ given in Eqs.~\eqref{IV4} and~\eqref{IV5}, respectively. That is, for $I=1,2$,
\begin{equation}\label{IV19}
\alpha_I ={\hat q}_I(0)= 4\pi \int_0^{\infty}r^2q_I(r)~dr=\alpha - (3-I)\frac{a_0}{\lambda_0}e^{\mu a_0}E_1(\mu a_0)\,,
\end{equation}
where $\alpha$ is given by Eq.~\eqref{IV8} and $E_1$ is the \emph{exponential integral function} given by~\cite{A+S}
\begin{equation}\label{IV20}
E_1(x):=\int_{x}^{\infty}\frac{e^{-t}}{t}dt\,.
\end{equation}
We recall that, for $x: 0 \to \infty$, $E_1(x)$ is a positive monotonically decreasing function  that behaves like $-\ln x$ near $x=0$ and decays exponentially as $x \to \infty$. It follows  that $0<  \alpha-\alpha_I \ll 1$ for sufficiently small $a_0/\lambda_0$, since  $0<a_0/\lambda_0 \ll 1$ and $0< \mu a_0 \ll 1$ (cf. Appendix A of Ref.~\cite{NL8}). Moreover, it follows from Eq.~\eqref{IV17d} that 
\begin{equation}\label{IV20a}
 \int K(x)~d^4x= \int \chi(\mathbf{x})~d^3x={\hat \chi}(0)\,,
\end{equation}
where ${\hat \chi}(0)$ is related to ${\hat q}(0)$ via Eq.~\eqref{IV2b}.

\subsection{Kernel $K$ of Linearized NLG}

The procedure followed above for the determination of kernel $R$ cannot be simply repeated for kernel $K$, since it turns out that the fundamental reciprocity relation~\eqref{III35a} could not be satisfied in this way. It is therefore necessary to determine $K$ via the Fourier transform method of Sec. III (cf. subsection A). Let us note that our basic expression for $R$ in Eq.~\eqref{IV16} implies that 
\begin{equation}\label{IV21}
{\hat R}(\xi)=\frac{\nu}{\nu -i\,\xi^0} \int e^{i\,\xi^0\,|\mathbf{x}|}\,q(\mathbf{x})\,e^{-i \boldsymbol{\xi} \cdot \mathbf{x}}~d^3x\,.
\end{equation}
Then ${\hat K}(\xi)$ is given by Eq.~\eqref{III40} and $K(x)$ can, in principle, be determined by inverse Fourier transformation. 

For a more tractable result, we can employ an approximation scheme that has already been introduced in Ref.~\cite{NL6} and involves neglecting certain retardation effects in Eq.~\eqref{IV16}. This means in practice that we
 replace\, $x^0-y^0 - |\mathbf{x}-\mathbf{y}|$ in Eq.~\eqref{IV16} by $x^0-y^0$; that is, instead of  Eq.~\eqref{IV16}, we consider 
\begin{equation}\label{IV22}
R(x-y)\approx \nu~e^{-\nu(x^0-y^0)}\, \Theta(x^0-y^0)\, q(\mathbf{x}-\mathbf{y})\,.
\end{equation}
The Fourier transform of this approximate kernel is
\begin{equation}\label{IV23}
\hat{R}(\xi) \approx \frac{\nu}{\nu-i\xi^0}~\hat{q} (|\boldsymbol{\xi}|)\,.
\end{equation}
If in Eq.~\eqref{IV22} we use for $q$ the spatial kernel $q_0$ given by Eq.~\eqref{IV7}, we get~\cite{NL7}
\begin{equation}\label{IV24}
\hat{q}_0(|\boldsymbol{\xi}|)=\frac{\mu}{\lambda_0(\mu^2+|\boldsymbol{\xi}|^2)}+\frac{1}{\lambda_0|\boldsymbol{\xi}|} \arctan{\Big(\frac{|\boldsymbol{\xi}|}{\mu}\Big)}\,. 
\end{equation}
We note that relation~\eqref{IV17b} is satisfied by both Eqs.~\eqref{IV21} and~\eqref{IV23}.

For Eq.~\eqref{IV23},  $1+\hat{R}\ne0$; hence, $K(x)$ can be obtained from 
\begin{equation}\label{IV25}
\hat{K} (\xi) \approx - \frac{\nu\, \hat{q} (|\boldsymbol{\xi}|)}{\nu\,[1+\hat{q} (|\boldsymbol{\xi}|)]-i\xi^0}\,.
\end{equation}
Let us note that in this case, Eq.~\eqref{IV17c} is satisfied. It can be shown, by means of contour integration and Jordan's Lemma, that~\cite{NL6}
\begin{equation}\label{IV26}
K(x) \approx -\frac{\nu}{(2\pi)^3}\Theta(x^0) \int \hat{q} (|\boldsymbol{\xi}|) e^{i \boldsymbol{\xi} \cdot \mathbf{x}}e^{-\nu(1+\hat{q})x^0}~d^3\xi\,.
\end{equation}
Moreover, it is straightforward to verify, by integrating this expression for $K(x)$ over all spacetime, that Eq.~\eqref{IV20a} is satisfied in this case.  Our approximation method has thus led to a manageable expression for kernel $K$; the nature and limitations of this simplification have been studied in Appendix C of Ref.~\cite{NL6}. 

Following the determination of the reciprocal kernel $R$ in Eq.~\eqref{IV16} and the approximate determination of kernel $K$, it is now possible to treat more explicitly the gravitational field of an isolated source in the linear post-Newtonian approximation of nonlocal gravity. We begin with the treatment of the  time-independent field of a stationary source in the next section, which amounts to a nonlocal extension of steady-state gravitoelectromagnetism (GEM) of  GR~\cite{M4}.  A dynamic nonlocal generalization of the standard GEM appears to be intractable.

\section{Gravitational Field of a Stationary Source}

The purpose of this section is to study the implications of the linearized nonlocal field equations in the \emph{transverse gauge} ($\overline{h}^{\mu\nu}{}_{, \nu}=0$)  for the weak time-independent gravitational field of an isolated \emph{stationary} source. To this end, let us note that in the field Eqs.~\eqref{III46}--\eqref{III47}, 
\begin{equation}\label{V1}
{\cal T}_{\mu \nu}(\mathbf{x}) =T_{\mu \nu}(\mathbf{x})+\int q(\mathbf{x}-\mathbf{y})\,T_{\mu \nu}(\mathbf{y})~d^3y\,,
\end{equation} 
as a result of Eq.~\eqref{IV17a}. In a similar way, we can show that ${\cal S}_{\mu \nu}=0$, since $S_{\mu \nu}=0$ in this case. To see this, let us consider Eq.~\eqref{III19} that defines $S_{\mu \nu}$; for a time-independent torsion field, Eq.~\eqref{III19} takes the form
\begin{equation}\label{V2}
\partial_i \int \chi(\mathbf{x}-\mathbf{y})\, \mathfrak{C}_{\mu}{}^{i}{}_{\nu}(\mathbf{y})~d^3y=-S_{\mu \nu} + \int \, \chi(\mathbf{x}-\mathbf{y})\,^{0}G_{\mu \nu}(\mathbf{y})~d^3y\,,   
\end{equation}
as a consequence of Eq.~\eqref{IV17d}. Following essentially the same steps as in our discussion of Eq.~\eqref{III19}, we find that $S_{\mu \nu}=0$, since the boundary surface in this case is at spatial infinity. Here, the seeming disappearance of the light cone is consistent with the complete temporal independence of the gravitational field. It follows from $S_{\mu \nu}=0$ and Eq.~\eqref{III24a} that the
 integral constraints in the stationary case reduce to $U_{[\mu \nu]}=0$, which contain only $\phi_{\mu \nu}$ and the constraints vanish for $\phi_{\mu \nu}=0$.  We can therefore set $\phi_{\mu \nu}=0$ in the gravitational potentials of a stationary source. In the transverse gauge, the linearized field equations~\eqref{III45a} of nonlocal gravity thus reduce in the stationary case to the 10 field equations
\begin{equation}\label{V3}
\nabla^2\,\overline{h}_{\mu \nu}(\mathbf{x}) = -2\kappa\, [T_{\mu \nu}(\mathbf{x})+\int q(\mathbf{x}-\mathbf{y})\,T_{\mu \nu}(\mathbf{y})~d^3y]\,.
\end{equation}
 
The spatial reciprocal kernel $q$ is independent of the speed of light; therefore, the standard static GEM approach can be adopted in this nonlocal case. Let us write the energy-stress tensor for a slowly rotating source with  $|\mathbf {v}| \ll c$ as $T^{00}=\rho c^2$ and $T^{0i}=c\,j^i$, where ${\bf j}=\rho\, {\bf v}$ is the matter current; moreover, the matter stresses are assumed to be independent of $c$ and of the form
$T_{ij}\sim \rho v_iv_j+P\delta_{ij}$, where $P$ is the pressure. Then, with $\overline{h}_{00}=-4\Phi /c^2$, we have a static gravitoelectric potential $\Phi(\mathbf{x})$ that satisfies Eq.~\eqref{IV3} of the Newtonian regime of nonlocal gravity. Next, $\overline{h}_{0i}=-2A_i/c^2$, where $\mathbf{A}(\mathbf{x})$ is the static gravitomagnetic vector potential that satisfies
\begin{equation}\label{V4}
\nabla^2\,\mathbf{A}(\mathbf{x}) = -\frac{8 \pi G}{c}\, [\,\mathbf{j}(\mathbf{x})+\int q(\mathbf{x}-\mathbf{y})\,\mathbf{j}(\mathbf{y})~d^3y]\,.
\end{equation}
It is interesting to note here the contribution of the ``dark" current, $\mathbf{j}_D(\mathbf{x})$, which is the convolution of the regular current with the reciprocal spatial kernel $q$, to the gravitomagnetic vector potential. The solution of Eq.~\eqref{V4} is thus given by
\begin{equation}\label{V4*} 
\frac{1}{2}\, \mathbf{A}(\mathbf{x})=\frac{G}{c}\int \frac{\mathbf{j}(\mathbf{y})+\mathbf{j}_D(\mathbf{y})}{|\mathbf{x}-\mathbf{y}|}~d^3y\,.
\end{equation}
Finally, Eq.~\eqref{V3} implies that $\overline{h}_{ij}=O(c^{-4})$ and is therefore neglected. Indeed, all terms of $O(c^{-4})$ are neglected in the standard linear GEM analysis~\cite{M4}.  

It is simple to check that the energy-momentum conservation law, Eq.~\eqref{III11a}, reduces in our nonlocal steady-state GEM treatment to $\boldsymbol{\nabla} \cdot \mathbf{j}=0$, which leads to $\boldsymbol{\nabla} \cdot \mathbf{j}_D=0$ as well, and is consistent with the transverse gauge condition $\boldsymbol{\nabla} \cdot \mathbf{A}=0$. With these conditions, one can develop a nonlocal version of the steady-state GEM for any suitable stationary source~\cite{PT}. In fact, with $ \mathbf {E}_g=\boldsymbol{\nabla} \Phi$ and $\mathbf{B}_g= \boldsymbol{\nabla} \times \mathbf{A}$, we have GEM fields with dimensions of acceleration such that
\begin{equation}\label{V4a} 
\boldsymbol{\nabla} \cdot \mathbf {E}_g=4\pi G\,(\,\rho+\rho_D)\,, \qquad \boldsymbol{\nabla} \times \mathbf{E}_g=0\,,
\end{equation}
\begin{equation}\label{V4b}   
\boldsymbol{\nabla} \cdot (\frac{1}{2} \mathbf{B}_g)=0\,, \qquad  \boldsymbol{\nabla} \times (\frac{1}{2}{\bf
B}_g)=\frac{4\pi G}{c}(\, \mathbf{j}+\mathbf{j}_D)\,.
\end{equation}
These are the steady-state field equations of nonlocal GEM. 

The GEM spacetime metric in this nonlocal case has the usual form~\cite{M4} 
\begin{equation}\label{V5} 
ds^2=-c^2\left(1+2\frac{\Phi}{c^2}\right)dt^2-\frac{4}{c}({\bf 
A}\cdot d{\bf
x})dt+\left(1-2\frac{\Phi}{c^2}\right) \delta_{ij}dx^idx^j\,.
\end{equation}
Here, $\Phi(\mathbf{x})$ is the gravitoelectric potential of nonlocal gravity in the Newtonian regime given by Eq.~\eqref{IV3} and $\mathbf{A}(\mathbf{x})=O(c^{-1})$ is the gravitomagnetic vector potential given by Eq.~\eqref{V4*}. It is now possible to discuss the motion of test particles and null rays that follow geodesics associated with this metric. For instance, for the motion of test particles, we recover the gravitational analog of the Lorentz force law~\cite{M4}.

In view of possible astrophysical applications, it is convenient to assume that the reciprocal kernel is  $q_0$ given by Eq.~\eqref{IV7}; then, $\Phi$ and $\mathbf{A}$ are given by
\begin{equation}\label{V6}
 \Phi(\mathbf{x})=-G\,\int \Big[1+\alpha (1- e^{-\mu r})+\frac{r}{\lambda_0}E_1(\mu r)\Big]\,\frac{\rho(\mathbf{y})}{|\mathbf{x}-\mathbf{y}|}~d^3y\,
\end{equation}
and 
\begin{equation}\label{V7}
 \frac{1}{2}\, \mathbf{A}(\mathbf{x})=\frac{G}{c}\,\int \Big[1+\alpha (1- e^{-\mu r})+\frac{r}{\lambda_0}E_1(\mu r)\Big]\,\frac{\mathbf{j}(\mathbf{y})}{|\mathbf{x}-\mathbf{y}|}~d^3y\,,
\end{equation}
where $r=|\mathbf{x}-\mathbf{y}|$ and $E_1$ is the exponential integral function defined in Eq.~\eqref{IV20}. Moreover, we note that~\cite{A+S}
\begin{equation}\label{V8}
\frac{\alpha}{2}\,\frac{\mu r}{\mu r +1}\,e^{-\mu r}<\frac{r}{\lambda_0}\, E_1(\mu r) \le \frac{\alpha}{2}\,e^{-\mu r}\,.
\end{equation}
These potentials can be explicitly calculated in any given situation involving an isolated material source using general methods familiar from classical electrodynamics~\cite{Ja}.
We are particularly interested in the propagation of light rays in this gravitational field. This is necessary in order to explain astrophysical phenomena associated with gravitational lensing without invoking dark matter. In linearized nonlocal gravity, just as in linearized GR, the effects due to gravitoelectric and gravitomagnetic fields could be treated separately  and then linearly superposed. Thus, as is well known, the bending of light rays due to the gravitoelectric potential $\Phi$ is given by twice the Newtonian expectation as worked out in detail in Ref.~\cite{NL8}. The influence of  the gravitomagnetic field on the propagation of light in GR  has been discussed in Refs.~\cite{M5, KM}. As explained in Ref.~\cite{M5}, according to GR, the gravitomagnetic bending of light rays passing near a slowly rotating source is generally smaller in magnitude than the gravitoelectric deflection by a factor of the order of $|\mathbf{v}|/c \ll 1$. It is therefore usually ignored in the discussion of gravitational lensing~\cite{SEF, PLW, VP}. The situation regarding the gravitomagnetic deflection of light in nonlocal gravity is, however, somewhat more complicated. For instance, if the integration in Eqs.~\eqref{V6} and~\eqref{V7} extends over a structure such as a cluster of galaxies for which $\mu r \gg 1$, then the quantity in square brackets in these equations essentially reduces to $1+\alpha$. Therefore, we are in effect working in the domain of linearized GR, but with enhanced gravity, i.e.,  with $G\to G(1+\alpha)$.

Imagine the propagation of light in the gravitational field of an isolated static source that moves uniformly with speed $c\, \beta$, $-1<\beta<1$ in the background Minkowski spacetime. This case is of interest in connection with the Bullet Cluster~\cite{BC1, BC2} and is treated in the next section; however, the general case of a time-dependent source is beyond the scope of this paper. 

\section{Light Deflection due to a Uniformly Moving Mass}

Consider the stationary case treated in Sec. V with no matter current.  In the rest frame of such a static gravitational source, it is convenient to think of this body in terms of a collection of fixed mass elements $m_j, j=0,1,2,...,N$. Then in Eq.~\eqref{V6}, we can write  
\begin{equation}\label{VI1}
 \rho(\mathbf{x})=\sum_jm_j\, \delta(\mathbf{x}-\mathbf{x}_j)\,, \qquad \Phi(\mathbf{x})=\sum_j m_j\, \varphi(|\mathbf{x}-\mathbf{x}_j|)\,,
\end{equation}
where,
\begin{equation}\label{VI2}
 \varphi(r)=-\frac{G}{r}\, \Big[1+\alpha (1- e^{-\mu r})+\frac{r}{\lambda_0}E_1(\mu r)\Big]\,.
\end{equation}
The spacetime metric in the rest frame of the source is given by Eq.~\eqref{V5} with $\mathbf{A}=0$. Let us remark here that for $\mu r \gg 1$, $\varphi(r) \approx -(1+\alpha)G/r$ in NLG, which is $1+\alpha$ times the Newtonian gravitational potential per unit mass. To return to GR, we can formally set $\lambda_0=\infty$ and $\alpha=0$ in NLG. 

In the background global inertial frame with coordinates $x^\mu=(t, x, y, z)$, the gravitational source under consideration here moves uniformly with speed $\beta$, $|\beta|<1$, along the $x$ axis. The source acts as a gravitational lens in deflecting a ray of light that, in its unperturbed state, is parallel to the $z$ axis, pierces the $(x,y)$ plane at the point $(a, b)$ and passes over the body.  We assume that the lens is relatively thin and its matter is mostly distributed in and near the $(x, y)$ plane. We are interested in the deflection of the ray by the lens when the point $(a, b)$ and the lens are in a definite geometric configuration as recorded by the static inertial observers at spatial infinity. It will turn out that the end result is independent of such a configuration. Let us assume that the desired configuration---i.e., the observationally preferred position of the source relative to the unperturbed ray of light---occurs at time $t=t_0$, when, for instance,  mass element $m_j$ of the lens is at $\mathbf{x}_j$. The source is then completely at rest in a comoving frame with coordinates  $x'^\mu=(t', x', y', z')$. To write the Lorentz transformation that connects the two frames, let us choose mass point $m_0$ to be the origin of the comoving system; then, 
\begin{eqnarray}\label{VI3}
\nonumber t'&=&\gamma[(t-t_0)-\beta(x-x_0)]\,, \\
\quad x'&=&\gamma[(x-x_0)-\beta(t-t_0)]\,, \quad y'=y-y_0\,, \quad z'=z-z_0\,.
\end{eqnarray}
Here, $\gamma$ is the Lorentz factor corresponding to $\beta$. Thus $m_0$ with coordinates $x^{\mu}_0=(t_0, x_0, y_0, z_0)$ is at the origin of coordinates in the rest frame of the source, namely, $x'^{\mu}_0=(0, 0, 0, 0)$. As the whole static source is at rest in the comoving frame at $t_0$, Eq.~\eqref{VI3} can be written with respect to any other mass point $m_j$ as 
\begin{eqnarray}\label{VI4}
\nonumber t'-t'_j&=&\gamma[(t-t_0)-\beta(x-x_j)]\,, \\
\quad x'-x'_j&=&\gamma[(x-x_j)-\beta(t-t_0)]\,, \quad y'-y'_j=y-y_j\,, \quad z'-z'_j=z-z_j\,,
\end{eqnarray}
where $t'_j=-\gamma \beta (x_j-x_0)$, etc. The result of the Lorentz transformation is that the invariant 
spacetime interval~\eqref{V5} can be written in the observers'  rest frame  as 
\begin{equation}\label{VI5}
 ds^2=(\eta_{\mu \nu} + h_{\mu \nu})\,dx^\mu\,dx^\nu\,,
\end{equation}
where the nonzero components of $h_{\mu \nu}$ are given by
\begin{equation}\label{VI6}
 h_{00}=h_{11}=-2\gamma^2(1+\beta^2)\,\Phi\,,
\end{equation}
\begin{equation}\label{VI7}
 h_{01}=h_{10}=4\beta \gamma^2\, \Phi\,, \qquad h_{22}=h_{33}=-2\,\Phi\,.
\end{equation}
Here, $\Phi$ depends upon time and is given by
\begin{equation}\label{VI8}
 \Phi=\sum_j m_j\, \varphi(u_j)\,,
\end{equation}
where $u_j=|\mathbf{x'}-\mathbf{x'}_j|$ is the \emph{positive} square root of 
\begin{equation}\label{VI9}
 u_j^2=\gamma^2\,[(x-x_j)-\beta(t-t_0)]^2+(y-y_j)^2 + (z-z_j)^2\,,
\end{equation}
in accordance with Eq.~\eqref{VI4}. In practice, $|\beta|\ll 1$; nevertheless, we perform the calculations in this section for arbitrary $\beta$, but then we set  $|\beta|\ll 1$ in the end result. To maintain our linear weak-field approximation scheme, however, $\beta^2$ cannot be too close to unity. Moreover, $\phi_{\mu \nu}=0$, and the transverse gauge condition is also maintained under Lorentz transformation. 

In the geometric optics approximation, a light ray propagates along a null geodesic
\begin{equation}\label{VI10}
\frac{d k^{\mu}}{d\lambda} + {^0}\Gamma^\mu_{\alpha \beta}\, k^\alpha k^\beta= 0\,,
\end{equation}
where the spacetime propagation vector $k^\mu=dx^{\mu}/d\lambda$ is tangent to the corresponding world line and $\lambda$ is an affine parameter along the path. Let $\tilde{k}^\mu = dx^{\mu}/d\tilde{\lambda}$ represent the unperturbed light ray whose trajectory is given by
\begin{equation}\label{VI11}
x(t)=a\,, \qquad y(t)=b\,, \qquad z(t)= \zeta + t-t_0\,,
\end{equation}
where $a$, $b$ and $\zeta$ are constants. To simplify matters in this case, we can choose $\tilde{\lambda}=t-t_0$, so that $\tilde{k}^\mu=(1, 0, 0, 1)$.

A comment is in order here regarding the physical significance of $\zeta$. In the regime of geometric optics, Eq.~\eqref{VI10} with $k^\mu=dx^{\mu}/d\lambda$ represents the equation of motion of the light particle (``photon") along the null ray. At $t=t_0$, $\zeta$ indicates the position of the unperturbed photon along the $z$ axis away from the $(x, y)$ plane. 

To calculate the deflection of light  from Eq.~\eqref{VI10}, we consider the net deviation $\Delta k^\mu$, 
\begin{equation}\label{VI12}
\Delta k^\mu= k^\mu(+\infty)-k^\mu(-\infty)=-\int_{-\infty}^{\infty}{^0}\Gamma^\mu_{\alpha \beta}\, k^\alpha k^\beta\, d\lambda\,,
\end{equation}
where $k^\mu(-\infty)=\tilde{k}^\mu$. The integrand here is computed along the null geodesic. To linear order, however, the calculation can be performed along the unperturbed light ray, namely, 
\begin{equation}\label{VI13}
\Delta k^\mu= -\int_{-\infty}^{\infty}{\mathcal L}^\mu(t_0+\tilde{\lambda}, a, b, \zeta+\tilde{\lambda})\, d\tilde{\lambda}\,,
\end{equation}
where $\tilde{\lambda}=t-t_0$ and
\begin{equation}\label{VI14}
{\mathcal L}^\mu(x)={^0}\Gamma^\mu_{\alpha \beta}(x)\, \tilde{k}^\alpha \tilde{k}^\beta\,.
\end{equation}
Here, the Christoffel symbols, 
\begin{equation}\label{VI15}
{^0}\Gamma^\mu_{\alpha \beta}= \frac{1}{2} \eta^{\mu \nu} (h_{\nu \alpha,\beta}+h_{\nu \beta,\alpha}-h_{\alpha \beta,\nu})\,,
\end{equation}
are determined from Eqs.~\eqref{VI6}--\eqref{VI9}. A detailed calculation reveals that ${\mathcal L}^\mu(t_0+\tilde{\lambda}, a, b, \zeta+\tilde{\lambda})$ can be expressed as
\begin{equation}\label{VI16}
{\mathcal L}^0=2 \gamma^2 \sum_j m_j\, \frac{1}{u_j}\frac{d\varphi(u_j)}{du_j}\,[\gamma^2 \tilde{\lambda}-\beta^3 \gamma^2(a-x_j)+(1+\beta^2)(\zeta-z_j)]\,,
\end{equation}
\begin{equation}\label{VI17}
{\mathcal L}^1=2 \gamma^2 \sum_j m_j\, \frac{1}{u_j}\frac{d\varphi(u_j)}{du_j}\,[\beta \gamma^2 \tilde{\lambda}+ (1-\beta^2 \gamma^2)(a-x_j)+2\beta (\zeta-z_j)]\,,
\end{equation}
\begin{equation}\label{VI18}
{\mathcal L}^2=2 \gamma^2 \sum_j m_j\, \frac{1}{u_j}\frac{d\varphi(u_j)}{du_j}\,(b-y_j)\,,
\end{equation}
\begin{equation}\label{VI19}
{\mathcal L}^3=2 \beta \gamma^2 \sum_j m_j\, \frac{1}{u_j}\frac{d\varphi(u_j)}{du_j}\,[(a-x_j)+\beta\,(\zeta-z_j)]\,.
\end{equation}
In principle, the integration in Eq.~\eqref{VI13} can now be carried through to determine the net deviation of the ray due to the gravitational attraction of the moving source; however, this calculation would involve
\begin{equation}\label{VI19a}
\frac{1}{r}\, \frac{d\varphi}{dr}=\frac{G}{r^3} \Big[1+\alpha-\alpha (1+\frac{1}{2}\mu r)e^{-\mu r}\Big]\,.
\end{equation}
We address the problem of calculating the relevant integrals in Appendix C. Using the results of Appendix C, we find that  for $\beta \ne 0$, 
\begin{equation}\label{VI20}
\Delta k^0=\beta \Delta k^1= \Delta k^3=-4 \beta \gamma\, G \sum_j \frac{m_j {\mathcal P}_j}{{\mathcal P}_{j}{}^2+ {\mathcal Q}_{j}{}^2}\Big[1+\alpha -\alpha\, \mathfrak{I}\big(\mu \sqrt{{\mathcal P}_j{}^2+{\mathcal Q}_j{}^2}\big)\Big]\,,
\end{equation}
\begin{equation}\label{VI21}
\Delta k^2=-4  \gamma\, G \sum_j \frac{m_j {\mathcal Q}_j}{{\mathcal P}_{j}{}^2+ {\mathcal Q}_{j}{}^2}\Big[1+\alpha -\alpha\, \mathfrak{I}\big(\mu \sqrt{{\mathcal P}_j{}^2+{\mathcal Q}_j{}^2}\big)\Big]\,,
\end{equation}
where
\begin{equation}\label{VI22}
 {\mathcal P}_j=(a-x_j)+\beta(\zeta-z_j)\,, \qquad  {\mathcal Q}_{j}=b-y_j\,.
\end{equation}
Moreover, $ \mathfrak{I}(x):= {\mathcal J}_2(x)+(x / 2){\mathcal J}_1(x)$, where ${\mathcal J}_1$ and  ${\mathcal J}_2$ are discussed in Appendix C; indeed,  
\begin{equation}\label{VI22a}
 \mathfrak{I}(x) = \int_0^{\infty} \frac{(1+\frac{1}{2}x\, \cosh{\upsilon})\,e^{-x\, \cosh{\upsilon}}}{\cosh^2{\upsilon}}~d\upsilon\,,
\end{equation}
so that $\mathfrak{I}(0)=1$ and $\mathfrak{I}(\infty)=0$. For $\alpha=0$, formulas~\eqref{VI20}--\eqref{VI22} extend the results of previous work on light deflection in GR~\cite{KM, KS, WS}.

With $z$ as the line-of-sight coordinate, the overall effect of the deflection of the light ray in the plane of the sky can be expressed via the angles
$\hat{\boldsymbol{\alpha}} = -( \Delta k^1,  \Delta k^2)$, where
\begin{equation}\label{VI23}
\hat{\boldsymbol{\alpha}}=4  \gamma \,G \sum_j m_j\, \frac{({\mathcal P}_j, {\mathcal Q}_j)}{{\mathcal P}_{j}{}^2+ {\mathcal Q}_{j}{}^2}\Big[1+\alpha -\alpha\, \mathfrak{I}\big(\mu \sqrt{{\mathcal P}_j{}^2+{\mathcal Q}_j{}^2}\big)\Big]\,.
\end{equation}
Other than an overall factor of $\gamma$, the effect of the motion of the gravitational source appears here in $\beta(\zeta-z_j)$ contained in ${\mathcal P}_j$. 

The end result for the deflection angle $\hat{\boldsymbol{\alpha}}$, and hence ${\mathcal P}_j$ and ${\mathcal Q}_j$, is independent of $t_0$ and any specific configuration of the lens and the photon. To illustrate this important point, we note that the photon crosses the $(x, y)$ plane at time $\bar{t}_0=t_0-\zeta$, when the point mass $m_j$, say, is at $(\bar{x}_j,  \bar{y}_j,  \bar{z}_j)$; then, repeating our calculation in this case would yield $ {\mathcal P}_j=(a-\bar{x}_j)-\beta \bar{z}_j$ and  ${\mathcal Q}_{j}=b-\bar{y}_j$. These are the same quantities as given in Eq.~\eqref{VI22}, since the lens has moved during the time interval $\zeta$; that is,  $x_j=\bar{x}_j + \beta \zeta$, $y_j=\bar{y}_j$ and $z_j=\bar{z}_j$.

Let us now suppose that the gravitational lens is thin---i.e., the extent of the deflecting mass in the $z$ direction is small~\cite{SEF}. Therefore, we may neglect $\beta z_j=\beta \bar{z}_j$ in ${\mathcal P}_j$, since in practice $|\beta| \ll 1$. Then, at the instant that the unperturbed photon crosses the lens plane, it is possible to express Eq.~\eqref{VI23} for a moving extended lens in a form that can be incorporated into the standard lens equation, namely, 
\begin{equation}\label{VI24}
\hat{\boldsymbol{\alpha}}(\boldsymbol{\theta})= \frac{4G}{c^2} \int \frac{\boldsymbol{\theta}-\overline{\boldsymbol{\theta}}}{|\boldsymbol{\theta}-\overline{\boldsymbol{\theta}}|^2}\Big[1+\alpha -\alpha\, \mathfrak{I}\big(\mu |\boldsymbol{\theta}-\overline{\boldsymbol{\theta}}| \big)\Big]\, \Sigma(\overline{\boldsymbol{\theta}})~d^2\overline{\theta}\,,
\end{equation}
where $\Sigma(\overline{\boldsymbol{\theta}})$ is the surface mass density of the deflecting source (``thin lens") and the integration is carried over the lens plane, which coincides with the $(x, y)$ plane. Thus,  in Eq.~\eqref{VI24},
\begin{equation}\label{VI25}
\boldsymbol{\theta}= (a, b)\,, \qquad  \overline{\boldsymbol{\theta}}=(\bar{x}, \bar{y})\,,
\end{equation}
where $\boldsymbol{\theta}$ is the unperturbed  position of the photon as it crosses the lens plane  and $\overline{\boldsymbol{\theta}}$ indicates the position of a point of the extended lens at that instant. 
Furthermore, it is possible to write
$\hat{\boldsymbol{\alpha}}=\boldsymbol{\nabla} \Psi$, where the lensing potential $\Psi$ is given by
\begin{equation}\label{VI26}
\Psi(\boldsymbol{\theta})= \frac{4G}{c^2} \int \big[\,\ln{|\boldsymbol{\theta}-\overline{\boldsymbol{\theta}}|}+\alpha\, \mathfrak{N}(\mu |\boldsymbol{\theta}-\overline{\boldsymbol{\theta}}|)\,\big]~\Sigma(\overline{\boldsymbol{\theta}})~d^2\overline{\theta}\,.
\end{equation}
Here, the first term in the integrand is the GR result, which follows from $\boldsymbol{\nabla} \ln |\mathbf{x}|= \mathbf{x}/|\mathbf{x}|^2$, while the nonlocal contribution to the lensing potential involves $\mathfrak{N}$, which is related to $\mathfrak{I}$ via $d\, \mathfrak{N}/d x = [1-\mathfrak{I}(x)]/x$.

It follows from these results that in the theoretical interpretation of  gravitational lensing data in accordance with nonlocal gravity, due account must be taken of the existence of the repulsive ``Yukawa" part of the gravitational potential as well. This may lead to the resolution of problems associated with light deflection by colliding clusters of galaxies. However, the confrontation of the theory with lensing data would require a separate detailed investigation that is beyond the scope of this work.

\section{Discussion}

This paper contains a new formulation of nonlocal gravity. Previous work on NLG~\cite{NL1, NL2, NL3, NL4, NL5, NL6, NL7, NL8} adopted the standpoint of gauge theories of gravitation, since GR$_{||}$, the teleparallel equivalent of general relativity that is rendered nonlocal in NLG via a constitutive ansatz, is indeed the gauge theory of the group of spacetime translations. In this approach to GR$_{||}$, the energy-momentum tensor $T_{\mu \nu}$ is not necessarily symmetric. There is, however, another way to approach GR$_{||}$, which is much closer to the spirit of GR. Within the Riemannian framework of GR, one can introduce a preferred tetrad frame and the associated Weitzenb\"ock connection; then, Einstein's gravitational field equations with an \emph{a priori} symmetric $T_{\mu \nu}$ can be formulated in terms of the Weitzenb\"ock torsion tensor. This is the approach that is adopted in the present paper. 

The distant parallelism of the preferred frame field can be viewed as a natural scaffolding on the spacetime manifold, reminiscent of the parallel frame field on Minkowski spacetime that would correspond to the parallel  tetrad frames of the static inertial observers at rest in a global inertial frame~\cite{M0}. It turns out that the nonlocal constitutive ansatz of the previous approach~\cite{NL1, NL2, NL3, NL4, NL5, NL6, NL7, NL8} must now be modified, since  the \emph{linearized} field equations of NLG with $T_{\mu \nu}=T_{\nu \mu}$ turn out to be inconsistent with the old ansatz. The general linear approximation of NLG with the new constitutive ansatz is then presented and the solutions of the linearized field equations are investigated. These new developments do not affect the main physical results of previous work~\cite{NL1, NL2, NL3, NL4, NL5, NL6, NL7, NL8} that consisted of the Newtonian regime of NLG and the treatment of linearized gravitational waves. In fact, our modification of the constitutive ansatz, which involves a constant overall parameter $p\ne 0$, primarily influences the gravitational field of time-varying sources in their near zones. All such complications disappear, however, for a stationary source. Indeed, it is possible to describe time-independent gravitational fields in terms of a simple gravitoelectromagnetic (GEM) metric familiar from GR. 

Nonlocality simulates dark matter. This important consequence of NLG is confirmed here in the linear approximation while preserving causality. With regard to possible astrophysical applications of linearized NLG to gravitational lensing, we consider the problem of deflection of light by a moving source. The results may be of interest in connection with gravitational lensing by merging clusters of galaxies.


\appendix{}
\section{Torsion and Contorsion}\label{appA}

The torsion tensor, defined in Eq.~\eqref{II9} in terms of the preferred frame field $e^\mu{}_{\hat{\alpha}}(x)$ has 24 independent components.  It is interesting to note that 
\begin{equation}\label{A1}
 \frac{1}{\sqrt{-g}}\,\frac{\partial}{\partial x^\mu}\, \Big(\sqrt{-g}\,e^\mu{}_{\hat{\alpha}} \Big)=-C_{\hat{\alpha}} \,,
 \end{equation}
 where the \emph{torsion vector} $C_\alpha$ is the trace of the torsion tensor. Moreover, it is possible to introduce a \emph{torsion pseudovector} $\check{C_\alpha}$ via the totally antisymmetric part of the torsion tensor $C_{[\alpha \beta \gamma]}$. Indeed, this axial vector is given by the dual of $C_{[\alpha \beta \gamma]}$, namely, 
\begin{equation}\label{A2}
\check{C_\alpha}=-\frac{1}{6} E_{\alpha \beta \gamma \delta}\, C^{[\beta \gamma \delta]}\,, \quad C_{[\alpha \beta \gamma]}=-E_{\alpha \beta \gamma \delta}\,\check{C^\delta}\,,
\end{equation}
where $E_{\alpha \beta \gamma \delta}=\sqrt{-g}\,\epsilon_{\alpha \beta \gamma \delta}$ is the Levi-Civita tensor and $\epsilon_{\alpha \beta \gamma \delta}$ is the alternating symbol with $\epsilon_{0123}=1$ in our convention. It is therefore possible to introduce a \emph{reduced torsion tensor} $T_{\alpha \beta \gamma} = - T_{\beta \alpha \gamma}$ with 16 independent components by subtracting out from $C_{\alpha \beta \gamma}$, in an appropriate fashion, its vector and pseudovector parts. In fact, the torsion tensor can be decomposed as 
\begin{equation}\label{A3}
C_{\alpha \beta \gamma}=-\frac{1}{3} (C_\alpha\, g_{\beta \gamma} -  C_\beta\, g_{\alpha \gamma})+ C_{[\alpha \beta \gamma]}+ T_{\alpha \beta \gamma}\,.
\end{equation}
It is straightforward to check from this \emph{definition} of the reduced torsion tensor that $T_{\alpha \beta \gamma}$ is totally traceless and $T_{[\alpha \beta \gamma]}=0$.

Similarly, from the definition of the contorsion tensor~\eqref{II12} as well as Eq.~\eqref{II17}, we find that 
\begin{equation}\label{A4}
K_{[\alpha \beta \gamma]}=\mathfrak{C}_{[\alpha \beta \gamma]}=-\frac{1}{2} C_{[\alpha \beta \gamma]}\,,
 \end{equation}
 \begin{equation}\label{A5}
K_{\alpha \beta \gamma}=-\frac{1}{3} (C_\beta\, g_{\alpha \gamma} -  C_\gamma\, g_{\alpha \beta})+ K_{[\alpha \beta \gamma]}+\frac{1}{2}(T_{\alpha \gamma \beta}+T_ {\beta \gamma \alpha}- T_{\alpha \beta \gamma})\,
\end{equation}
and
 \begin{equation}\label{A5a}
\mathfrak{C}_{\alpha \beta \gamma}=\frac{2}{3} (C_\alpha\, g_{\beta \gamma} -  C_\beta\, g_{\alpha \gamma})+ \mathfrak{C}_{[\alpha \beta \gamma]}+\frac{1}{2}(T_{\alpha \beta \gamma}+T_ {\alpha \gamma \beta}- T_{\beta \gamma \alpha})\,.
\end{equation}
Let us note here the following useful formulas
\begin{equation}\label{A6}
g^{\mu \nu}  K_{\mu \nu}{}^\sigma=C^\sigma\,, \qquad g^{\mu \nu}  \mathfrak{C}_{\sigma \mu \nu}:=-\mathfrak{C}_\sigma=2\,C_\sigma\,,
\end{equation}
\begin{equation}\label{A7}
 \Gamma^{\alpha}_{\beta \alpha}=\,{^0} \Gamma^\alpha_{\beta \alpha} = \frac{1}{\sqrt{-g}}\,\frac{\partial}{\partial x^\beta}\, (\sqrt{-g}\,)\,, \qquad  \Gamma^{\alpha}_{\alpha \beta}= \Gamma^{\alpha}_{\beta \alpha}+C_\beta\,,
\end{equation}
\begin{equation}\label{A8}
g^{\mu \nu}\, \Gamma^{\alpha}_{\mu \nu}=-C^\alpha- \frac{1}{\sqrt{-g}}\,\frac{\partial}{\partial x^\beta}\, \Big(\sqrt{-g}g^{\alpha \beta}\Big)\,, 
\end{equation}
\begin{equation}\label{A9}
K_{\alpha}{}^{\mu \nu}K_{\mu \nu \beta}=-K_{\alpha}{}^{\mu \nu}\,\Gamma^\gamma_{\mu \nu}\,g_{\gamma \beta} =-\frac{1}{2}K_{\alpha}{}^{\mu \nu} C_{\mu \nu \beta}\,
\end{equation}
and $\nabla_\gamma\,g^{\alpha \beta}=0$, which can be written as 
\begin{equation}\label{A10}
 g^{\alpha \beta}{}_{, \gamma}= -\Gamma^\alpha_{\gamma \delta}\, g^{\delta \beta} - \Gamma^\beta_{\gamma \delta}\, g^{\delta \alpha}\,.
\end{equation}

\section{Constitutive Relation of NLG}\label{appB}

This appendix is devoted to a discussion of the constitutive relation of nonlocal gravity. More precisely, we wish to examine the \emph{local} connection between $X_{\mu \nu \rho}$ and the torsion tensor in Eq.~\eqref{II26} and its implications for linearized NLG. Ultimately, of course, the confrontation of the theory with observation can determine the right relation. 

Imagine, for instance, the possibility of choosing $X_{\mu \nu \rho}=\mathfrak{C}_{[\mu \nu \rho]}$. Returning to the general form of the linearized field Eqs.~\eqref{III18a}--\eqref{III18b}, we have in this case
\begin{equation}\label{B1}
X_{(\mu}{}^{\sigma}{}_{\nu)} = 0\,, \qquad X_{[\mu}{}^{\sigma}{}_{\nu]}=\frac{1}{2}\,\eta^{\sigma \rho}\,\phi_{[\mu \rho , \nu]}\,,
\end{equation}
since in the linear approximation $\mathfrak{C}_{[\mu \rho \nu]}=\frac{1}{2}\,\phi_{[\mu \rho\,, \nu]}$. Thus Eq.~\eqref{III18a} is the same here as in the linearized Einstein equation of GR and Eq.~\eqref{III18b} takes the form
\begin{equation}\label{B2}
\eta^{\sigma \rho}~\partial_\sigma\,\int K(x-y)\, \phi_{[\mu \rho , \nu]}(y)~d^4y=0\,.
\end{equation}
In this case, we have a \emph{complete} separation of the 10 dynamic metric variables $\overline{h}_{\mu \nu}$ from the 6 tetrad variables $\phi_{\mu \nu}$. The integral constraints~\eqref{B2} can be satisfied with
\begin{equation}\label{B3}
\phi_{\mu\nu}=0\,. 
\end{equation}
Thus at the linear level, this theory of nonlocal gravity is essentially equivalent to local GR; therefore, the connection between nonlocal gravity and dark matter disappears in this case.  

In connection with the separation of the metric variables from the tetrad variables, let us consider the possibility that 
\begin{equation}\label{B4}
X_{\mu \nu \rho}= \mathfrak{C}_{\mu \nu \rho}+\frac{1}{2}\,\mathfrak{C}_{\rho \mu \nu}\,.
\end{equation}
It is useful to note that we now have in Eqs.~\eqref{III18a}--\eqref{III18b}, 
\begin{equation}\label{B5}
X_{(\mu}{}^{\sigma}{}_{\nu)} = \mathfrak{C}_{(\mu}{}^{\sigma}{}_{\nu)}\,, \qquad X_{[\mu}{}^{\sigma}{}_{\nu]}=\frac{3}{4}\,\eta^{\sigma \rho}\,\phi_{[\mu \rho , \nu]}\,.
\end{equation}
The constraint equations in this case contain the secondary tetrad variables $\phi_{\mu \nu}$ exclusively. Thus to simplify matters, one can again assume that $\phi_{\mu \nu}=0$; then, the constraint equations are satisfied and the ten dynamic nonlocal field equations depend solely upon $\overline{h}_{\mu \nu}$. However, we note that in this case
$X_{\mu \nu \rho} \ne -X_{\nu \mu \rho}$, so that ${\cal N}_{\mu \nu}$ in Eq.~\eqref{II29a} does not in general transform as a tensor under arbitrary coordinate transformations. Thus this case violates the basic geometric structure of nonlocal gravity theory. 

Clearly, one can concoct other combinations and study their consequences; however, the rest of this appendix is devoted to a detailed discussion of the difficulty associated with the simplest possibility, namely, $X_{\mu \nu \rho}= \mathfrak{C}_{\mu \nu \rho}$, adopted, along with the possibility that $T_{\mu \nu}\ne T_{\nu \mu}$,   in previous work on this subject~\cite{NL1, NL2, NL3, NL4, NL5, NL6, NL7, NL8}. In the present work, 
$T_{\mu \nu}= T_{\nu \mu}$ as in GR; however, $X_{\mu \nu \rho}= \mathfrak{C}_{\mu \nu \rho}$ then leads, in a manner that is independent of any gauge condition, to a contradiction. 
The field equations in this case can be obtained from Eqs.~\eqref{III18c}--\eqref{III25} for $p=0$, and we recall here that $S_{0\mu}=0$. Let us take 
\begin{equation}\label{B6}
 ^{0}G_{00}= \kappa\,  {\mathcal T}_{00}\,             
\end{equation}
from the set of field equations for the metric variables and write it using Eq.~\eqref{III8} as
\begin{equation}\label{B7}
\overline{h}_{00, i}{}^i-\overline{h}_{ij,}{}^{ij}=-2 \kappa\,  {\mathcal T}_{00}\,,
\end{equation} 
where ${\mathcal T}_{00}$ is the total energy density of the source defined by Eq.~\eqref{III41}. Next, we take Eq.~\eqref{III24b} from the set of integral constraint equations, namely, 
\begin{equation}\label{B8}
\int K(x-y)\,\delta (x^0-y^0- |\mathbf{x}-\mathbf{y}|){\mathcal W}_i(y)~d^4y=0\,,
\end{equation} 
where, in agreement with Eq.~\eqref{III24aA}, ${\mathcal W}_i $ is given by
\begin{equation}\label{B9}
{\mathcal W}_i =-\phi_{ij,}{}^{j} -\big(\overline{h}_{00, i}-\overline{h}_{ij,}{}^{j}\big)\,.
\end{equation} 
Integrating over the temporal coordinate in Eq.~\eqref{B8}, we find
\begin{equation}\label{B10}
\int K(|\mathbf{x}-\mathbf{y}|, \mathbf{x}-\mathbf{y})\, {\mathcal W}_i(x^0-|\mathbf{x}-\mathbf{y}|, \mathbf{y})~d^3y=0\,.
\end{equation} 
We note that 
\begin{equation}\label{B11}
\delta^{ik}\,{\mathcal W}_{i,k} =-\overline{h}_{00, i}{}^i+\overline{h}_{ij,}{}^{ij}\,,
\end{equation} 
since $\phi_{ij}=-\phi_{ji}$. Hence, we find from Eq.~\eqref{B7} the interesting result that 
\begin{equation}\label{B12}
\delta^{ij}\,{\mathcal W}_{i,j} =2 \kappa\,  {\mathcal T}_{00}\,.
\end{equation}  

To demonstrate that Eq.~\eqref{B12} is in general incompatible with Eq.~\eqref{B10}, we apply the partial derivative operator $\partial/\partial x^j$ to Eq.~\eqref{B10}. To simplify the calculation, let us define the functions $\eta$ and $F$ by
\begin{equation}\label{B13}
\eta:=x^0-|\mathbf{x}-\mathbf{y}|\,, \qquad F(\mathbf{x}-\mathbf{y}) := K(|\mathbf{x}-\mathbf{y}|, \mathbf{x}-\mathbf{y})\,.
\end{equation} 
Then, we have that 
\begin{equation}\label{B14}
 \frac{\partial \eta}{\partial x^j} = -\frac{\partial \eta}{\partial y^j}\,, \qquad \frac{\partial F}{\partial x^j} = -\frac{\partial F}{\partial y^j}\,.
\end{equation} 
Hence, taking the derivative of Eq.~\eqref{B10} results in
\begin{equation}\label{B15}
\partial_j \int F\, {\mathcal W}_i~d^3y=\int \Big[-\frac{\partial F}{\partial y^j}\,{\mathcal W}_{i}(\eta, \mathbf{y})+ F \frac{\partial \eta}{\partial x^j}\,{\mathcal W}_{i,0}(\eta, \mathbf{y})\Big]~d^3y=0\,.
\end{equation} 
Using integration by parts, we find that
\begin{equation}\label{B16}
 \int \frac{\partial}{\partial y^j}\,(F {\mathcal W}_i)~d^3y=\int  F\Big[\frac{\partial}{\partial y^j}\,{\mathcal W}_{i}(\eta, \mathbf{y})+ \frac{\partial \eta}{\partial x^j}\,{\mathcal W}_{i,0}(\eta, \mathbf{y})\Big]~d^3y\,.
\end{equation} 
From 
\begin{equation}\label{B17}
\frac{\partial}{\partial y^j}\,{\mathcal W}_{i}(\eta, \mathbf{y})= \frac{\partial \eta}{\partial y^j}\,{\mathcal W}_{i,0}(\eta, \mathbf{y}) +{\mathcal W}_{i,j}(\eta, \mathbf{y})\,
\end{equation} 
and Eq.~\eqref{B14}, we see that in Eq.~\eqref{B16} terms involving ${\mathcal W}_{i,0}$ cancel; thus,  Eq.~\eqref{B16} can be written as 
\begin{equation}\label{B18}
 \int \frac{\partial}{\partial y^j}\,(F {\mathcal W}_i)~d^3y=\int  F\,{\mathcal W}_{i,j}~d^3y\,.
\end{equation} 
Taking the trace of this equation and using Gauss's theorem, we finally get from Eq.~\eqref{B12} that 
\begin{equation}\label{B19}
\int  F\,\Big(\delta^{ij}\,{\mathcal W}_{i,j}\Big)~d^3y=2 \kappa\, \int K(|\mathbf{x}-\mathbf{y}|, \mathbf{x}-\mathbf{y})\,{\mathcal T}_{00}(\eta, \mathbf{y})~d^3y=0\,.
\end{equation} 
This important result can also be expressed as
\begin{equation}\label{B20}
 \int  W(x-y) T_{00}(y)~d^4y=0\,,
\end{equation} 
where kernel $W$ is given by Eq.~\eqref{III43d}. 

The source of the gravitational field has been assumed to be finite and isolated in space, but is otherwise arbitrary. It is conceivable that Eq.~\eqref{B20} could be satisfied for rather special source configurations. In general, however, Eq.~\eqref{B20} is not satisfied for an arbitrary source, which indicates that a solution of the field equations does not exist. We have thus shown, without using any gauge condition, that the metric part of the field equations of NLG is in general incompatible with the tetrad part for $X_{\mu \nu \rho}= \mathfrak{C}_{\mu \nu \rho}$. The incompatibility proof can be directly extended to constitutive relations of the forms $X_{\mu \nu \rho}= \mathfrak{C}_{\mu \nu \rho}+ p' \,\mathfrak{C}_{[\mu \nu \rho]}$ and $X_{\mu \nu \rho}= \mathfrak{C}_{\mu \nu \rho}+ p'' \,E_{\mu \nu \rho \sigma}\,C^{\sigma}$, where $p' \ne 0$ and $p'' \ne 0$ are constant parameters.

Let us now consider the constitutive relation adopted in the present paper. Then, instead of Eq.~\eqref{B6}, we have 
\begin{equation}\label{B21}
 ^{0}G_{00}= \kappa\,  {\mathcal T}_{00}-p\, {\mathcal U}_{00}\,,             
\end{equation}
where
\begin{equation}\label{B22}
 U_{00}=\int K(x-y) \check{C}_{0,0}(y)~d^4y\,, \qquad  {\mathcal U}_{00}=-\int R(x-y) \check{C}_{0,0}(y)~d^4y\,             
\end{equation}
and we have used here the reciprocity relation~\eqref{III35a}. It follows from Eqs.~\eqref{III8} and~\eqref{B11} that 
\begin{equation}\label{B23}
\delta^{ij}\,{\mathcal W}_{i,j} =2 \kappa\,{\mathcal T}_{00}+2\,p\int R(x-y)\check{C}_{0,0}(y)~d^4y\,.
\end{equation}  
Next, the relevant integral constraint is in this case $S_{[i\,0]}=p\,U_{[i\,0]}$, or
\begin{equation}\label{B24}
\int K(|\mathbf{x}-\mathbf{y}|, \mathbf{x}-\mathbf{y})\, {\mathcal W}_i(x^0-|\mathbf{x}-\mathbf{y}|, \mathbf{y})~d^3y=4\,p\,U_{[i\,0]}\,.
\end{equation} 
Hence, using the approach adopted above for the $p=0$ case, we have
\begin{equation}\label{B25}
\int K(|\mathbf{x}-\mathbf{y}|, \mathbf{x}-\mathbf{y})\,\Big(\delta^{ij}\,{\mathcal W}_{i,j}\Big)(x^0-|\mathbf{x}-\mathbf{y}|, \mathbf{y})~d^3y=4\,p\,\delta^{ij}\partial_j\,U_{[i\,0]}\,.
\end{equation} 
It follows from Eq.~\eqref{B23} that 
\begin{equation}\label{B26}
 \kappa\, \int K_c(x-y)\,{\mathcal T}_{00}(y)~d^4y+\,p\int \int K_c(x-z)R(z-y)\check{C}_{0,0}(y)~d^4y~d^4z=2\,p\,\delta^{ij}\partial_j\,U_{[i\,0]}\,,
\end{equation}  
where $K_c$ is defined by Eq.~\eqref{III24d}. Calculating $U_{[i\,0]}$ from Eq.~\eqref{III23c} and using $\check{C}^{\sigma}{}_{,\sigma}=0$, we find 
\begin{equation}\label{B27}
\delta^{ij} \partial_j \int K(|\mathbf{x}-\mathbf{y}|, \mathbf{x}-\mathbf{y})\,\check{C}_i(\eta, \mathbf{y})~d^3y=\int K_c(x-y)\check{C}_{0,0}(y)~d^4y\,.
\end{equation}   
Moreover,  
\begin{equation}\label{B28}
\delta^{ij} \partial_j \int K(x-y)\,\check{C}_{[i,0]}(y)~d^4y=\frac{1}{2}\partial_{\sigma} \int K(x-y)(\check{C}^{\sigma}{}_{,0}-\check{C}_{0,}{}^{\sigma})(y)~d^4y\,,
\end{equation}   
which, after using Gauss's theorem and $\check{C}^{\sigma}{}_{,\sigma}=0$, results in
\begin{equation}\label{B29}
\delta^{ij} \partial_j \int K(x-y)\,\check{C}_{[i,0]}(y)~d^4y=-\frac{1}{2}\int K(x-y)(\square\, \check{C}_{0})(y)~d^4y\,.
\end{equation}   
Putting all these results together and using the definition of kernel $W$ in Eq.~\eqref{III43d}, we finally arrive at a nonlocal integral constraint for  $\check{C}_{0}$,
\begin{equation}\label{B30}
\kappa \int W(x-y)\,T_{00}(y)~d^4y=-p \int [\,W(x-y)\,\check{C}_{0,0}(y)+ K(x-y)\,\square\, \check{C}_{0}(y)\,]~d^4y\,.
\end{equation}    
We assume that this equation for $\check{C}_0$ can be solved---for example, via Fourier analysis---in terms of $T_{00}$, the energy density of the gravitational source. In this way, for $p\ne 0$, we avoid the contradiction that has forced us to introduce the additional term in the constitutive relation of this work. 

\section{Light Deflection Integrals}\label{appC}

In Eqs.~\eqref{VI16}--\eqref{VI19} of Sec. VI, consider
\begin{equation}\label{C0}
\frac{1}{r}\,\frac{d\varphi}{dr}=\frac{G(1+\alpha)}{r^3} -\alpha G\Big(1+\frac{1}{2}\,\mu r\Big)\,\frac{e^{-\mu r}}{r^3}\,,
\end{equation}
where the first part on the right-hand side is simply due to Newtonian attraction augmented by $1+\alpha$, while the second repulsive ``Yukawa" part is due to the requirements of nonlocality. To compute the net deflection of light, the integrals due to the first part of Eq.~\eqref{C0} are simpler and we therefore treat them first. 

Let $w(X)>0$ be given by 
\begin{equation}\label{C1}
w(X)={\mathcal A} +2\,{\mathcal B}\,X+{\mathcal C}\,X^2\,,          
\end{equation} 
where $\tilde{\Delta}:={\mathcal A}\,{\mathcal C}-{\mathcal B}^2 \ne 0$.
It is then straightforward to verify that 
\begin{equation}\label{C2}
\int \frac{dX}{w^{3/2}}=\frac{{\mathcal B}+{\mathcal C}X}{\tilde{\Delta}\,w^{1/2}}\,,  \qquad  \int \frac{X\,dX}{w^{3/2}}=-\frac{{\mathcal A}+{\mathcal B}X}{\tilde{\Delta}\,w^{1/2}}\,,      
\end{equation} 
where only positive square roots are considered throughout. Let us now assume that ${\mathcal C}>0$ and $\tilde{\Delta}>0$, so that 
\begin{equation}\label{C2a}
{\mathcal C}\,w(X)=({\mathcal C}\,X +{\mathcal B})^2+\tilde{\Delta}\,.          
\end{equation} 
Hence, $w>0$ for $X: -\infty \to +\infty$.  In this case, we have 
\begin{equation}\label{C3}
{\mathcal I}_1=\int_{-\infty}^{\infty} \frac{dX}{w^{3/2}}= \frac{2\,{\mathcal C}^{1/2}}{\tilde{\Delta}}\,, \qquad   {\mathcal I}_2=\int_{-\infty}^{\infty} \frac{X\,dX}{w^{3/2}}=- \frac{2\,{\mathcal B}}{\tilde{\Delta}\,{\mathcal C}^{1/2}}\,.       
\end{equation}

For the problem of light deflection discussed in Sec. VI, we have $w(t-t_0)=u_{j}^2$, where $u_j$ is given by Eq.~\eqref{VI9}. That is, along the unperturbed ray,
\begin{equation}\label{C4}
u_j^2= {\mathcal A}_j +2\,{\mathcal B}_j\,(t-t_0)+{\mathcal C}_j\,(t-t_0)^2\,,         
\end{equation}
where 
\begin{equation}\label{C5}
{\mathcal A}_j=\gamma^2(a-x_j)^2+(b-y_j)^2+(\zeta-z_j)^2\,, \qquad {\mathcal B}_j=-\beta \gamma^2(a-x_j)+(\zeta-z_j)\,         
\end{equation}
and ${\mathcal C}_j=\gamma^2$. Moreover, we find that $\tilde{\Delta}_j={\mathcal A}_j\,{\mathcal C}_j-{\mathcal B}_j{}^2 = \gamma^2 ({\mathcal P}_j{}^2+{\mathcal Q}_j{}^2)$, where ${\mathcal P}_j$ and ${\mathcal Q}_j$ are defined in Eq.~\eqref{VI22} and $\tilde{\Delta}_j$, by assumption, never vanishes. Thus the conditions for the applicability of Eq.~\eqref{C3} are satisfied and with $X=t-t_0$, we find that the integrals for the first part are given by
\begin{equation}\label{C6}
{\mathcal I}_1=\int_{-\infty}^{\infty} \frac{dX}{u_j^3}= \frac{2\,\gamma^{-1}}{{\mathcal P}_j{}^2+{\mathcal Q}_j{}^2}\,, \qquad   {\mathcal I}_2=\int_{-\infty}^{\infty} \frac{X\,dX}{u_j^3}= \frac{2\,\gamma^{-1}[\,\beta\, {\mathcal P}_j-(\zeta-z_j)]}{{\mathcal P}_j{}^2+{\mathcal Q}_j{}^2}\,,       
\end{equation}
which, together with the results given below for the second part of Eq.~\eqref{C0}, eventually lead to Eqs.~\eqref{VI20}--\eqref{VI21} of Sec. VI.

To treat the integration of the second (``Yukawa") part of Eq.~\eqref{C0}, let us first note that 
\begin{equation}\label{C7}
u_j= \big(\hat{u}_j^2 + \hat{\Delta}_j^2\big)^{1/2}\,,         
\end{equation}
where 
\begin{equation}\label{C8}
\hat{u}_j=\gamma X +\gamma^{-1}\, {\mathcal B}_j\,, \qquad   \hat{\Delta}_j=({\mathcal P}_j{}^2+{\mathcal Q}_j{}^2)^{1/2}\,.         
\end{equation} 
As $X:-\infty \to +\infty$,  $\hat{u}_j$ also goes from $-\infty$ to $+\infty$; therefore, it proves useful to introduce a new variable $\upsilon: -\infty \to +\infty$ such that 
\begin{equation}\label{C9}
\hat{u}_j= \hat{\Delta}_j\, \sinh{\upsilon}\,, \qquad u_j= \hat{\Delta}_j\, \cosh{\upsilon}\,.         
\end{equation}
The calculation of the integrals for the second part then ultimately reduces to the determination of ${\mathcal J}_1(\vartheta_j)$ and ${\mathcal J}_2(\vartheta_j)$, where
\begin{equation}\label{C10}
\vartheta_j:=\mu\,  \hat{\Delta}_j >0\,        
\end{equation}
and
\begin{equation}\label{C11}
{\mathcal J}_n(\vartheta):= \int_0^{\infty} \frac{e^{-\vartheta\, \cosh{\upsilon}}}{\cosh^n{\upsilon}}~d\upsilon\, 
\end{equation}
for  $n=1,2,3,...$.
It is interesting to observe that  ${\mathcal J}_n(0)=(\sqrt{\pi}/2)\Gamma(\frac{n}{2})/\Gamma(\frac{n+1}{2})$ and ${\mathcal J}_n(\infty)=0$.

To determine ${\mathcal J}_1$ and ${\mathcal J}_2$, let us first note that ${\mathcal J}_1(0) =\pi/2$ and ${\mathcal J}_2(0)=1$. Moreover, for $0<|\epsilon|\ll 1$, we find from Eq.~\eqref{C11} that for $\vartheta >0$,
\begin{equation}\label{C12}
{\mathcal J}_1(\vartheta + \epsilon)={\mathcal J}_1(\vartheta) -\epsilon K_0(\vartheta)+...\,,
\end{equation}
\begin{equation}\label{C13}
{\mathcal J}_2(\vartheta + \epsilon)={\mathcal J}_2(\vartheta) -\epsilon {\mathcal J}_1(\vartheta)+\frac{1}{2} \epsilon^2 K_0(\vartheta)+...\,,
\end{equation}
where $K_0(\vartheta)$ is the modified Bessel function given by~\cite{A+S}
\begin{equation}\label{C14}
K_0(\vartheta)= \int_0^{\infty} e^{-\vartheta\, \cosh{\upsilon}}~d\upsilon\,.        
\end{equation}
For $x: 0 \to \infty$, $K_0(x)$ behaves as $-\ln{x}$ for $x\to 0$, but then rapidly decreases monotonically with increasing $x$ and vanishes exponentially as $x \to \infty$. In fact, 
\begin{equation}\label{C15}
K_0(x) \sim \sqrt{\frac{\pi}{2x}} \, e^{-x}\,     
\end{equation}
for $ x \to \infty$~\cite{A+S}.  It follows from Eqs.~\eqref{C12}--\eqref{C13} that 
\begin{equation}\label{C16}
\frac{d {\mathcal J}_1}{d \vartheta}=-K_0(\vartheta)\,, \qquad  \frac{d {\mathcal J}_2}{d \vartheta}=- {\mathcal J}_1(\vartheta)\,.     
\end{equation}
Therefore, the series expansion for $K_0$~\cite{A+S} can be employed to find  ${\mathcal J}_1$
\begin{equation}\label{C16}
{\mathcal J}_1(\vartheta)=\frac{\pi}{2}-\int_0^{\vartheta}K_0(x)~dx\,,
\end{equation}
which in turn will help determine ${\mathcal J}_2$ via
\begin{equation}\label{C17}
{\mathcal J}_2(\vartheta)=1-\int_0^{\vartheta}{\mathcal J}_1(x)~dx\,.
\end{equation}
In practice, the polynomial approximation for $K_0$~\cite{A+S} can be used to develop  corresponding polynomial approximations for ${\mathcal J}_1$ and ${\mathcal J}_2$.

\end{document}